# Cancer genomics, targeted therapy to precision oncology: the treatment paradigm is a holistic approach to cancer therapeutics

**Manish Kumar**
School of Bio Sciences & Technology
Vellore Institute of Technology, Vellore, Tamil Nadu, India

**Email:** kmanish125@yahoo.com, kmanish.bt125@gmail.com
**ORCID:** https://orcid.org/0000-0002-6021-7063

**CORRESPONDENCE**
1.
C/O. Sri M. Narayan Swami
7/1A,1st Street, Bhavani Nagar
Katpadi P.O. Dharapadavedu,
Vellore- 632007, Tamil Nadu, India

2.
S/O. Nityanand Prasad
ME-010-0017, Dhruvneeti Colony
Near Samarpan Hospital, Rajneeti Road
Madhopur, P.O. Basudeopur
Munger- 811202, Bihar, India

**Abstract:** Cancer is a fatal genetic disease involving unregulated cell growth and proliferation. Cellular behaviors such as growth, proliferation, survival, or death rely upon the ability of cells to communicate with the surrounding system; therefore, effective targeting of dysregulated signaling pathways involving growth factors, regulatory proteins, cell adhesion molecules, and immune system molecules is critical to cancer therapy. They are mainly driven by alterations in tumor suppressor genes and oncogenes that may vary among different cancer types. Importantly, patients with the same cancer type respond differently to available cancer treatments, likely due to tumor-specific DNA, RNA, and proteins, indicating the need for patient-specific treatment options. Precision oncology has evolved as a form of cancer therapy focused on genetic and molecular profiling of tumors to identify specific molecular alterations involved in carcinogenesis for tailored individualized cancer treatment. This article aims to briefly explain the foundations and frontiers of precision oncology, highlighting the cutting-edge innovations in tools and techniques associated with it to assess its scope and importance in achieving efficacious cancer treatment over time.

## 1. Introduction

Cancer is a devastating disease that causes one in six deaths globally and has considerable physical, psychological, and economic impacts on people affected by the disease. It continues to be the second most common cause of hospital death after heart disease, most of which can be prevented by early diagnosis and improved prevention and treatment strategies for the disease. Techniques for the efficient diagnosis of cancer accompanied by the development of efficacious treatment options, and a better understanding of the socioeconomic factors that affect cancer incidence, prevalence, and related deaths across the globe are needed [1,2]. More than 100 cancer types with subtypes have been identified on the basis of location, cell of origin, and genetic variations that influence cancer development and therapeutic response. Most cancers appear in epithelial cells as carcinomas, such as lung, skin, breast, liver, colon, prostate, and pancreas cancer, whereas sarcomas arise from mesenchymal tissues, originating in myocytes, adipocytes, fibroblasts, and osteoblasts. Tumors also develop frequently in hematopoietic tissues, such as leukemia and lymphomas, and in nervous tissues, e.g., gliomas and neuroblastomas. They are among the most common types of cancer, taking a high toll in terms of life and property throughout the world [3,4]. Thus, considering the vast number of cancer cases worldwide, a formal initiative towards fighting the menace of cancer first appeared in the United States in the form of the National Cancer Act of 1971 and was signed by President Richard Nixon to promote cancer research and the application of outcomes for minimizing cancer incidence and mortality rates associated with the disease. The act was euphemistically described as the "War on Cancer", and the National Cancer Program that was borne from this initiative resulted in a concerted effort to develop the infrastructures across the length and breadth of the country for the treatment, cure, and eradication of cancer [5]. A similar approach was adopted by most other developed and developing nations in the following years to combat the deadly disease, which has succeeded in satisfying the purpose involved to a good extent since then despite the facts and figure that suggest demographic factors play a role in cancer management [6,7]. Overall morbidity from cancer has decreased, and net survival rates, both short-term and long-term, have increased substantially for all cancers combined in recent decades. The survival rates for cancer types that are responsive to therapy surpass 90% in developed countries, and the prognosis for several other cancer types that were considered the deadliest diseases earlier has improved noticeably in recent years owing to the rapid advances realized in clinical oncology over the years. [8,9]. However, the fight against cancer is far from complete, as an estimation by the World Health Organization (WHO) in 2018 revealed that the incidence of cancer is expected to double to approximately 37 million new cases by 2040, with no confirmed remedies for most cancer types [10,11]. While researchers continue their endeavors to identify the exact causes of different cancer types and subtypes and develop strategies for prevention, diagnosis, and treatment, cancer remains the leading cause of death worldwide and has a major impact on societies across the globe. Many types of cancer therapies are currently available, such as chemotherapy, immunotherapy, hormonal therapy, targeted drug therapy, radiation therapy, surgery, and stem cell transplantation. One may receive a single type of treatment or a combination of therapies, but regardless of the treatment regimen, a much-needed cure for many cancers remains largely elusive [12]. Therefore, a holistic approach to cancer treatment that effectively addresses the complexities of disease progression, therapeutic resistance and recurrence is needed.

Advances in the fields of cellular and molecular biology, genetic engineering and biotechnology including recent developments in computational techniques and drug development must address the problem at a fairly convincing level over time.

## 2. Genetic and biochemical basis of cancer development

The tumor is an abnormal mass of tissue that appears due to unregulated growth in the division of cells, which successfully prevents senescence. A tumor is benign until it is limited to its original position and becomes malignant or cancerous when it is capable of growing and metastasizing to other parts of the body. Rigorous research in the past few decades, supported by advances in cell and molecular biology, has led scientists to clearly understand that genetic changes associated with cancer incidence cause the disease to grow and spread to other parts of the body. Cancer is initiated as a result of uncontrolled cell division and proliferation, leading to tumor formation, which results in metastasis involving the dissemination of cancer cells from the original or primary tumor through the circulation of blood or lymph, and invasion of other normal tissues and organs to form secondary tumors at distant locations in the body, which is responsible for approximately 90% of cancer- related deathsreported globally.Cell proliferation requires a balanced rate of cell growth and division to maintain an increase in cell numbers for growth and development, maintenance of tissue homoeostasis and wound healing. The fundamental abnormality leading to cancer development is unwanted cell proliferation due to an absence of balance between cell division and cell loss through cell death and differentiation. Cell division relies on cell cycle regulation, which generally involves extracellular growth-regulatory signals as well as internal signaling proteins that monitor the genetic integrity of the cell to ascertain that cellular development progresses well in time. It depends on progression through distinct phases of the cell cycle and is regulated by several cyclin-dependent kinases (CDKs) that act in association with their cyclin partners. Alterations in the overall expression pattern of cyclins cause the cellular process go awry and proliferate rapidly, resulting in tumor formation. Most of the related events accompanying tumor formation and cancer progression, such as cell differentiation, apoptosis, angiogenesis, invasion and metastasis, are guided similarly by alterations in the expression patterns of regulatory portions owing to changes (mutations) in the genes of interest, and the factors that cause these changes often tend to provoke cancer development [13]. Genetic mutations can be inherited or acquired mutations that appear later in life. Acquired mutations are of somatic origin, are much more common and cause most cancers. As the somatic mutation theory (SMT) is evidence-based, it has become the dominant theory in cancer research. In fact, cancer is a multistep process involving the initiation and progression of random mutations in certain key genes, such as oncogenes or tumor suppressor genes, which lead to the manifestation of cancer. Every single gene in the body is most likely to have undergone deleterious changes or mutations in its DNA sequence on a number of occasions in the cell's lifetime, whereas the repair mechanism in place would restrict noticeable changes. In this way, the generation of cancer must be conclusively linked to sustained gene mutations caused by either external agents called mutagens, which often lead to the appearance of different somatic variants, or certain critical changes that might have been inherited in the body. Importantly, a single mutation will not be enough to transform a normal cell into a cancer cell, as it would require a number of changes to accumulate in the cells in due course for cancer development to occur. For example, mutations in the most pronounced cancer-causing genes, such as RAS (derived from rat sarcomavirus) or MYC (derived from myelocytoma, a cancer of the myelocytes), may not lead to unchecked proliferation until changes in repressor genes, such as RB and TP53 which encode components of protective mechanisms have not occurred simultaneously. Thus, multiple genetic changes are typically required for the development of cancer, so it must be seen as an evolutionary process involving both genetic changes and selection [14]. Multiple rate- limiting steps can work against the development of cancer, with persistent changes accelerating the process. Thus, most cancers are thought to be derived from a single abnormal cell or a small group of cells with a few deleterious gene mutations followed by the accumulation of additional changes in some of their descendants, allowing them to outgrow others in number and resulting in tumorous growth in the body. Moreover, cancer can also be driven by epigenetic changes that alter the gene expression pattern of cells without accompanying alterations in the DNA sequence of the cell [15]. Some physical modifications in the chromatin structure that are capable of influencing the pattern of gene expression are often led by DNA methylation, histone modifications, and miRNA-based alterations inside the cell. Epigenetic regulation of DNA and RNA usually controls how genes are turned on or off and thus plays important roles in maintaining normal cell behavior, whose deregulation causes alterations in gene expression patterns to potentially influence tumorigenesis. These changes are frequently accompanied by sustained exposure of the affected cells to several stressful external stimuli presented by certain environmental factors and/or lifestyle-related changes that may involve nutrition, toxicants, alcohol, etc. Although epigenetic changes do not alter the sequence of DNA, the process might cause point mutations and disable DNA repair mechanisms frequently involved in cancer development. Traditionally, epigenetic and genetic changes have been seen as two separate mechanisms that independently participate in carcinogenesis, which may not be the only possible mechanism involved in cancer development. Recent studies from whole-exome sequencing (WES), the technique for sequencing all of the protein-coding regions of genes in a genome, for thousands of human cancers have revealed the presence of many inactivating mutations in genes that can potentially disrupt DNA methylation patterns, histone modifications, and nucleosome positioning and hence control the epigenome to contribute to cancer progression. Thus, both the genome and epigenome can regulate the progression of cancer through associated mutations. Therefore, interference between the two is highly anticipated and can be exploited to provide new possibilities for cancer treatment [16].

Cancer ultimately remains a selective multistep process triggered by mutations leading to the activation of specific oncogenic pathways with the concurrent inactivation of tumor suppressor genes that act as sentinels to control unwanted cell growth and proliferation. Scientists have been trying to analyze the totality of cancer-causing gene mutations, which are regarded as the “mutational landscape” of different types of cancer, and to target them effectively for cancer cure. In fact, most of these biochemical processes are conserved in model organisms, such as the free-living transparent nematode Caenorhabditis elegans and the fruit fly Drosophila melanogaster, along with other large animal models, and are widely used for ease of genetic manipulation to study the complex biology of cancer. Somatic cell mutations, called somatic structural variants (SVs), have been shown to account for more than half of all cancer-causing mutations. These variants or mutations differ from the hereditary or germline variants that have passed from parents to offspring and become incorporated into the DNA of every cell in the body. These SVs can be observed in transformed cells and in their daughter cells, which may continue to grow because of errors in DNA copying and their repair mechanisms during cell division, thereby altering the genomic structure, which becomes more numerous with time. Although somatic SVs play crucial roles in cancer development, relatively little is known about their mode of action in cancer development. Methods to detect and identify the functional effects of these SVs are sure to enable researchers to understand the molecular consequences of individual somatic mutations in cancer. The findings related to mutation-specific molecular alterations could be used to develop therapies that target mutated cells, opening great possibilities in cancer therapy [17].

Furthermore, most of the human genome consists of noncoding regions, and studies on variations in the noncoding regions of cancer cells reveal additional mechanisms underlying cancer progression. For example, changes in noncoding regions such as point mutations and complex genomic rearrangements can disrupt or create transcription factor- binding sites or even affect noncoding RNA loci, leaving options for unwanted changes in the gene expression pattern of the cell. Cancer whole-genome sequencing (WGS) remains the most comprehensive method for identifying variants in noncoding regions, as targeted approaches such as WES may miss certain variants residing outside coding regions. Pieces of evidence suggest that oncogenesis typically involves interplay between germline and somatic variants, and different modes of action of noncoding variants could further potentiate these developments. Thus, a systematic approach to unravel the roles of the noncoding genome in cancer progression should help improve cancer diagnosis and therapy [18].

An important aspect of cancer biology is that all cellular behaviors are manifestation of underlying cellular physiology and biochemistry that are ultimately guided by enzymes whose timely availability is controlled by genes. Enzymes catalyze specific reactions within compartments of the cells to maintain a balanced state of tissues and organs. Cancer-based genomic studies highlight the many ways in which enzyme activities can be altered to contribute to cancer development owing to certain genetic changes. Importantly, kinetic parameters associated with enzymatic activities are tangibly altered to influence cancer initiation and progression. Therefore, enzyme-based studies of cancer cells can provide critical insight into the molecular and biochemical mechanisms of cancer progression and help determine the effectiveness of anticancer agents, mechanisms of treatment resistance and disease relapse [19]. Additionally, changes in the tumor microenvironment (TME) can critically affect enzymatic activities and exaggerate cancer development. Enzymes specifically linked to the regulation of key cellular behaviors such as cell proliferation, death and differentiation may have a direct influence on cancer development, but some enzymes required for many other activities may also be involved in the incidence of cancer because of their crucial role in maintaining tissue homeostasis. For example, monoamine oxidase A (MAOA) is a mitochondrial enzyme found in animal tissues that catalyzes the breakdown of biogenic monoamines, and is commonly known for its ability to regulate neurotransmitters such as dopamine, adrenaline, and serotonin. As cells of the nervous system and immune system have many common surface receptors and secretory molecules, they may share many common cellular pathways crucial to health and disease. The evidence suggests that MAOA is involved in other diseases, including cancer, cardiovascular disease, and diabetes, in addition to its role in neurobiology. MAOA can inhibit the activities of different types of tumor-associated immune cells, such as T cells and macrophages, and has been implicated in the regulation of antitumor immune responses. MAOA inhibitors are being studied for their potential in combination therapy to improve the effectiveness of cancer immunotherapy [20].

Moreover, epidemiological studies have consistently shown that environmental factors or lifestyle changes involving mutagenic agents are the primary culprits. Thus, it is necessary not only to associate genetic mutations with different cancers but also to work on the mechanism of action of mutagens by focusing on enzymes that invariably mediate oncogenic transformations. For example, overexpression of the enzyme ribonucleotide reductase (RnR), which catalyzes the formation of deoxyribonucleotides from ribonucleotides necessary for cell division, is implicated in many forms of cancer, and the genes encoding the components of the enzyme are often mutated, leading to hyperactivity of the enzyme. However, there are instances indicating that cytoplasmic material rather than the karyoplast is mainly responsible for cellular transformation, which might be better explained as a consequence of certain external influences, including epigenetic modulations, than purely genetic changes [21]. RnR active site inhibitors have been developed to biophysically deactivate the enzyme, when necessary, with positive outcomes.

Furthermore, the tumor microenvironment (TME) is an integral part of tumors and plays a central role in all stages of cancer progression. Importantly, the activation and remodeling of stromal cells at the origin of cancer development precedes the formation of metastases. Each organ has a unique microenvironment where resident stromal cells are considered essential for tissue integrity and repair. Stromal cells refer to connective tissue cells that are heterogeneous in nature and form the structural framework of organs. These cells surround parenchymal cells, the organ-specific cells, and support their activities crucial for different biological processes, such as maintaining tissue homeostasis, wound healing, and immune responses, which ultimately define the primary function of an organ. Stromal cells mainly consist of fibroblasts, macrophages, self-renewing and multipotent mesenchymal stromal cells also known as mesenchymal stem cells (MSCs), immune cells, endothelial cells, and components of the basement membrane. These cells are thought to be sentinels of tissue integrity as many of the cells in the stroma possess tumor-suppressing capabilities, but their transition to being dysfunctional modulators of angiogenesis and metastasis is common in cancer progression [22]. The tumor stroma may secrete growth factors, cytokines, extracellular matrix (ECM) proteins and many other regulatory proteins that are thought to promote cell growth, survival and migration to promote the metastatic spread of cancer cells. As an important component of the tumor microenvironment, tumor ECM differs considerably from that of normal tissues. ECM provides essential signals to maintain tissue architecture, polarity, and regulate cell growth and apoptosis. It therefore plays a critical role in tumor formation, angiogenesis and metastasis. ECM gene mutations have been implicated in many tumors, including breast, ovarian, prostate, lung, pancreatic, colon cancer, hepatoma and melanoma. Stromal cells also contribute to the regulation of ECM remodeling and the release of the core matrisome, ECM protein signatures, and certain bioactive ECM fragments called matrikines or matricryptins that may play a critical role in controlling cancer progression. Therefore, a better understanding of the complexity of interactions of stromal cells with cancer cells and other components of the TME seems necessary to design effective therapeutic options to prevent metastatic development and diseases relapse [23, 24].

## 3. Cellular basis of cancer progression and metastasis

Certain disruptions in the physiological balance between cell proliferation and cell death prolong cell survival and proliferation and are thought to be important steps in carcinogenesis. The transition from a non-malignant cell to a malignant cell is governed by mechanisms similar to those involved in normal cell growth and differentiation. Growing epithelial cells can gradually lose their polarity and adhesion contacts to become invasive cancer cells, a process often accompanied by an epithelial-mesenchymal transition (EMT). Cancers of different tissues utilize somewhat different patterns to ultimately converge to a common path of cancer development in the form of tumor growth followed by angiogenesis, invasion, and metastasis. All such developments are ultimately guided by genetic and epigenetic changes associated with cancer cells and supported by certain tissue-specific factors that enable the tissue to exploit these changes to meet its specific needs, resulting in reprogramming of the molecular events utilized by different cancer cells, and no gene change is thought to be common to all cancers. Because uncontrolled cell growth and proliferation remain the most evident causes of cancer, certain alterations in the pattern of cell death and differentiation promoting overall cell survival could further aggravate the gradual transformation of tissue from normal to tumorous and from benign to metastatic. As expected, observations confirm that evasion of cell death by apoptosis and autophagy is the hallmark property of most, if not all, cancers and actively contributes to cell growth and proliferation. Apoptosis, the process of programmed cell death, also known as type 1 cell death, is mediated through caspase degradation activated by mitochondria. It is employed for removing damaged cells and is crucial to the early development and overall maintenance of tissue homeostasis. Loss of apoptotic control enables cancer cells to survive longer, allowing more time for the accumulation of mutations, which can deregulate cell proliferation and differentiation and stimulate angiogenesis and metastasis. Autophagy is the major intracellular degradation system mediated by lysosomes and involves the engulfment of unwanted proteins and damaged organelles in double-membraned vesicles called autophagosomes for destruction and recycling. Autophagy can play a protective role in promoting cell survival, but excessive autophagy plays a suppressive role by inducing autophagic cell death, known as type 2 cell death. Autophagy is universally accepted to play a tumor-suppressive role at the early stage, whereas defective autophagy is associated with tumorigenesis. Deregulation of these essential catabolic pathways contributes to the development of a tumor and is often involved in promoting invasion and metastasis. Cancer cells can develop novel mechanisms for evading apoptosis and autophagy, and new discoveries have revealed the possible interaction between these two catabolic pathways. The evidence suggests that the inhibition of apoptosis causes autophagy, whereas autophagy inhibition induces apoptosis. These findings may help the key proteins and intermediates involved in these pathways be exploited successfully in cancer therapeutics [25]. In addition, the ability of cancer cells to maintain constant proliferative capacity may be guided by their transformation into persistent non-senescent cells. In this context, telomeres are specific repeating DNA structures found at the ends of the chromosome of the cell that protect the genome against unnecessary nucleolytic degradation, recombination, repair, and interchromosomal interactions. Telomeres are maintained by telomerase, which adds nucleotides to telomeres to prevent them from becoming shorter. Germ cells typically express high levels of telomerase to maintain telomere length. In somatic cells, telomere length usually decreases over time, leading cells to undergo senescence with age. Loss of cells in this way generally acts as a barrier to tumor growth, and the transformed cells escape as they maintain their telomeres despite repeated cell divisions because these cells are able to express high levels of active telomerase. Telomerase has become a potential target in cancer therapeutics because it is overexpressed in transformed cancer cells and cancer stem cells in diverse forms of malignancies. Telomere maintenance mechanisms (TMMs) are used by cancer cells through telomerase activation and sometimes by alternate means called alternative lengthening of telomeres (ALTs) to avoid apoptosis. Anti- telomerase therapeutics have been developed to selectively target cancer cells to induce cell death via apoptosis without affecting normal cells [26].

An important feature of cancer is that the population of cells that make up cancer is profoundly heterogeneous at the genetic and epigenetic levels. Tumors usually represent a heterogeneous mass of distinctly differentiated cells that include connective tissue cells, immune cells, cancer stem cells, and vasculature, and these subpopulations of cells can be further distinguished by a variety of features impacting their phenotype that generally involve genetic alterations. Tumors develop this feature mainly because the cancer genome is unstable due to the accumulation of many cancer-causing gene mutations. Genomic instability further promotes genetic diversity by providing the raw material for the generation of tumor heterogeneity [27]. Importantly, there are fragile points in every genome where the DNA is more likely to be mutated when the genome is replicated. These breakage points have frequently been linked to genetic and heritable disorders such as cancer. Moreover, there can be mutations present in certain genes, known as mutator mutations, that further increase the inherent rate of genomic changes, resulting in even greater genetic instability that leads to the accumulation of multiple oncogenic mutations within a cellular lineage. Not all such changes are "malignant", but the rate of such development could translate into cancer manifestation at different stages in a lifetime. Mutator mutations and genetic instability are generalized concepts in cancer genetics, referred to as the mutator hypothesis, which relates to those few mutations that lead to an increased rate of gene mutations leading to chromosomal instability, microsatellite instability, and deregulation of activities related to DNA damage and repair [28]. In addition, there are transposable elements (TEs) present in cells called 'jumping genes', which are repetitive sequences of DNA that move from one place to other in the genome by different means and represent almost half of the human genome. They represent a powerful means of genetic modification and have played an important role in the evolution of genomes. TEs are typically regulated from the beginning, at the early stage of development and throughout the lifespan mainly by epigenetic mechanisms such as DNA methylation and histone modifications and are crucial for maintaining genomic stability through the regulation of the transcriptomic and proteomic profiles of the cell. Dysregulation of TEs has been implicated in different types of human cancers, with the possibility of chromosomal aberrations, oncogenic activation, transcriptional dysregulation, and noncoding RNA aberrations as potential mechanisms underlying the development of cancer [29]. Moreover, the gradual accumulation of oxidative damage to critical biomolecules such as DNA due to persistent metabolic oxidative stress and inflammation also contributes to genomic instability and related diseases, including cancer, indicating relevant measures for prevention and treatment. This feature of cancer cells has also guided researchers to kill vulnerable cells by inducing lethal genomic instability in the cells through radiation therapy and chemotherapy. It is a rather nonselective means of killing cancer cells with associated side effects, which could be improved by devising methods to selectively target the affected cells inside the body. Researchers have begun examining the genomic data of vulnerable individuals to allow clinicians to embark on personalized radiation therapy [30].

A crucial component of tissue heterogeneity found in tumors, known to be responsible for drug resistance and recurrence, is cancer stem cells (CSCs), which are at the forefront of cancer research owing to their potential to induce cancer development. Recent studies have shown that different subpopulations of CSCs within the tumor mass can be identified on the basis of the expression of cancer stem cell surface markers on normal stem cells with characteristics similar to those of normal stem cells, such as self-renewal and multilineage differentiation capabilities, with a much longer half-life than that of most other cells. The intrinsic properties of self-renewal, multipotency, and longevity render stem cells more susceptible to accumulating gene mutations, leading to neoplastic transformation, as proposed by the cancer stem cell hypothesis [31,32]. They have been found to be the key drivers of tumorigenicity, tumor heterogeneity, recurrence, and drug resistance in many cancer types, and different targeted molecules, including nanoparticle-based drug delivery systems, are being tested for effectively targeting CSC-related pathways for cancer treatment [33.34]. Moreover, the immune cells in the tumor mass can differ greatly, and an emerging finding of tumor heterogeneity is that tumors from different patients have different degrees of immune cell infiltration and immune cell compositions. The immunologically "hot" tumors present elevated levels of T-cell infiltration, so these tumors are more susceptible to immunotherapy than immunologically "cold" tumors that do not allow similar T-cell infiltration. This immunogenic heterogeneity simply impacts treatment outcomes and may direct treatment planning [35,36].

## 4. Cancer genomics and targeted cancer therapy to precision oncology

Changes in genes involved in cell growth, proliferation, differentiation, or death appeared to be essential to the overall change in cell behavior; therefore, cancer needed to be viewed as a genetic disease and treated accordingly for better outcomes. The realization that cancer progression required the participation of cellular genes provided one of the important rationales for undertaking the Human Genome Project (HGP) in 1986. The HGP project in fact revolutionized cancer care by providing the foundational reference sequence required to identify cancer-driving mutations, shifting oncology from treating localized tumors to targeting specific genetic alterations. Since its completion in 2003, the human genome sequencing has affirmed its pledge of being an essential resource for understanding cancer pathogenesis. Technological advances in the field of molecular biology have been exploited over the years to unravel genomic changes to fully understand the pathogenesis of human cancer. The range of cancer-causing mutations is known to be very large, and the mutational landscape differs from one another depending on the type of cancer; even people suffering from the same cancer type are found to have considerably different mutation patterns. Moreover, it has long been known that every patient responds differently to particular treatments despite having the same type and stage of cancer. These observations have been compelling and led researchers to adopt a precision medicine approach to cancer therapy, necessitating the study of the genetic features of vulnerable individuals for a patient-specific treatment regimen towards the most effective treatment of cancer. Since the nineteenth century, biometricians have been interested in decoding the relationship between genetics and diseases and attempting to understand the roles of "constitutional" and "environmental factors" in the distribution of diseases. Werner Kalow's 1962 textbook 'Pharmacogenetics' published on the issue of heredity and the response to drugs, emphasizing the importance of relating the response of therapeutic drugs to their biochemistry and the role of genetics and evolution in shaping individual-level differences. Advances in genetic engineering and the consequent understanding of clinically relevant genetic variations over the years have revolutionized how a range of diseases can be diagnosed and treated in the clinic, exploiting the genetic peculiarities of individuals, and the idea needs to be adequately applied to cancer research for better outcomes. Consequently, in past decades, precision medicine approach to cancer treatment has emerged as a field of cancer research taking into account the genetic specificities of individuals for efficient cancer treatment. The term precision oncology has been coined for specific clinical oncology practices that rely upon genomic profiling of individual tumors for complete molecular characterization of transformed cells and tissues to identify and target specific molecular alterations for efficient cancer therapy [37]. The emergence of next-generation sequencing (NGS) in 2005, has proven to be highly important in this direction. Massive parallel sequencing or high throughput sequencing (HTS) is another term for next-generation sequencing (NGS), as this technology allows the sequencing of multiple samples in a single run, increasing the speed and efficiency of sequencing compared to traditional sequencing methods. The ability to process large amounts of genetic information quickly has helped to gain a better understanding of how our genes work and how diseases develop. This technology can now be efficiently used to determine the order of nucleotides in entire genomes or selected regions of DNA or RNA to study genetic variation associated with different biological processes or diseases which is critical for precision oncology [38]. Tumor and cell-free DNA profiling, proteomic and RNA-based analyses, as well as markers of the immune system can be used to identify characteristics features of cancer towards designing anticancer therapy regimen well suited for individual patients. Precision oncology ultimately aims to achieve perfectly planned cancer therapy by designing custom-tailored treatment for individuals based on the identification of their unique needs for the best possible results. Innovative clinical trials have evolved shifting focus from tumor type to tumor-agnostic, molecular profiling tests with the aim to improve treatment outcomes. A number of precision medicine clinical trials have been conducted. The majority of these trials confirm that targeted matched therapy is associated with superior results compared to non-matched therapy therapy across tumor types. Recent precision oncology trials supported by the NCI and other agencies, such as the NCI- MATCH, also known as MATCH (Molecular Analysis for Therapy Choice), the NCI-MPACT (Molecular Profiling-based Assignment of Cancer Therapy), the ALCHEMIST (Adjuvant Lung Cancer Enrichment Marker Identification and Sequencing Trial), the TAPUR (Targeted Agent and Profiling Utilization Registry), and the DRUP, (Drug Rediscovery Protocol), have significantly helped shift the focus from cancer treatment on type and origin to target tumor-specific genetic variations for better outcomes [39]. Trastuzumab is a humanized monoclonal antibody that received FDA approval in 1998 as a targeted cancer drug for HER2-positive cancers. This drug has proven its effectiveness over time and is considered a cornerstone of precision oncology for HER2-positive cancers, such as breast and gastric cancers, improving survival through targeted inhibition of tumor growth. It can be used as adjuvant therapy or as first-line treatment for metastatic disease. The discovery and approval of imatinib as the first signal transduction inhibitor(STI), for the treatment of acute myeloid leukemia in 2002 virtually established targeted cancer therapy as a means to an end and marked the beginning of the precision oncology approach to cancer therapy. The use of precision oncology in clinics thus began approximately 25 years ago, but it has significantly improved the effectiveness of cancer treatment over time and is about to enter mainstream clinical practices [40]. This approach should be used in the early stages of cancer management, and vulnerable people should have easy access to tumor profiling and effective personalized therapy to ensure early cancer detection and treatment results.

## 5. Selective targeting of genetic alterations in medical oncology

Traditionally, cancer treatments such as chemotherapy and radiation therapy have targeted actively growing cells in the tissue instead of just attacking diseased cells, resulting in a variety of side effects. Therefore, a deeper understanding of the molecular events underlying cancer progression was realized decades ago to develop treatments that selectively target affected cells to alleviate the serious side effects of cancer treatment. The functional roles of many critical players involved in tumor growth, tissue invasion, and metastasis have been described precisely in recent decades on the basis of the draft of the human genome and other related developments that took place in the following years [41]. As noted previously, advances in DNA sequencing have revealed that cancer genomes can exhibit thousands of somatic genetic alterations, such as point mutations, DNA copy number variations, differences in RNA transcription and protein expression and epigenetic changes. The findings also revealed many crucial genes and proteins associated with cancer reprogramming pathways, which could be attractive targets for precise cancer treatments.

Furthermore, tumors of similar types and conditions can possess unusually heterogeneous mutation patterns; however, these mutations may actually influence the same characteristic cellular pathways and networks. The molecules generally involved are thought to participate in crucial cellular events in different ways, eventually leading to uncontrolled cell growth and proliferation, which are responsible for tumor growth. However, it remains generally unclear which mutational changes could be considered the primary drivers of disease progression or how these changes influence cancer pathophysiology, and extensive genomic data mining and orthogonal modeling are needed to gain insight into the biological mechanisms underlying different cancers. An overview of some of the frequently mutated genes and their protein products in different cancers may be useful for better understanding how certain proteins, with their specific roles in cellular processes, including interactions and networking with many different molecules of the cell upstream or downstream of the chain, may play essential roles in cancer cell reprogramming and disease progression, as well as their importance in cancer treatment. A few common alterations that are frequently implicated in cancer progression with profound effects are detailed below.

**Retinoblastoma (RB) and tumor protein p53 (TP53):** The RB and TP53 are central tumor suppressors that play crucial roles in cell cycle regulation and genome integrity and are frequently altered in various forms of human cancers. The RB tumor suppressor gene that encodes the Rb protein is a master regulator of the cell cycle that is often mutated or functionally inactivated during cancer development. Rb proteinsform complexes with the E2F family of transcription factors and thereby repress or downregulate several genes that encode key regulators of cell progression through the cell cycle, apoptosis, and DNA repair to preserve genome stability. Their transcriptional repression by the Rb-E2F complex can be relieved through the phosphorylation of RB, leading to committed cell cycle progression, which can be reversed again at the level of cyclin-dependent kinases. The gene product can also interact with chromatin remodelers and modifiers to repress certain genes crucial for cell cycle progression. TP53 encodes tumor protein p53 (TP53), a 53 kDa weighted nuclear protein that plays a crucial role in regulating the cell cycle and apoptosis and thereby controls cell division and cell death.

This protein functions primarily to ensure normal cell growth and proliferation and is also responsible for maintaining genome stability. It is the key player in the tumor suppressive DNA damage response (DDR). ATM (ataxia-telangiectasia mutated), ATR (ATM- and Rad3-related), and other related protein kinases are the initial DDR kinases that help p53 sense damage to DNA and activate other genes to repair damage or suppress cell division to prevent the accumulation of oncogenic mutations that often lead to tumor development. This task is supported by p21, a cyclin-dependent kinase inhibitor (CKI) activated by p53, which serves as a cell cycle inhibitor and anti-proliferative effector of the cell. Stresses such as viral infection or DNA damage, a relatively common oncogenic act, turn on p53 functions, leading to cell cycle arrest for DNA repair, senescence for permanent growth arrest, or apoptosis for programmed cell death. Mutations in the p53 gene not only disable their tumor suppressive function but can also engage in cancer-promoting activities by gaining oncogenic properties or inactivating the remaining suppressive elements in the cell. Therefore, mutations in the gene cause cancer cells to grow and spread throughout the body. A wide variety of mutations have also been identified in the p53 gene, which often occur late during cancer progression. An estimated 40–50% of human cancers carry deleterious mutations in the regulatory p53 gene [42].

**Myelocytomatosis protein (MYC):** Myc proteins are potent tumor inducing proteins derived from MYC oncogenes, a group of related proto-oncogenes, and are commonly involved in the pathophysiology of human cancer. MYC alone may not cause transformative effects, and studies have revealed that changes in the expression pattern of tumor suppressors such asTP53 alongside MYC synergistically induce proliferation, survival, and metastasis. It is also a known target of RB repressor deregulation, which may result in increased MYC activity and consequent tumorigenic effects. The Myc proto-oncogene family has three members, C-MYC, MYCN, and MYCL, which encode the transcription factors c-Myc, N-Myc, and L-Myc, respectively. They are essential transcription factors involved in the activation of many protein-coding genes associated with many different biological processes, including cell proliferation and differentiation, cell metabolism, cell cycle progression, apoptosis, and self-renewal of stem cells. Myc oncoproteins have been shown to mandate tumor cell fate by inducing stemness and blocking differentiation and cellular senescence, or irreversible cell cycle arrest which contributes to cancer progression. Additionally, MYC can influence changes in the tumor microenvironment to induce angiogenesis and/or suppress the host immune response. The c-Myc oncoprotein forms a crucial part of a dynamic cellular network whose members interact selectively with one another and with many of the transcriptional coregulators and histone-modifying enzymes that support the maintenance of sustained cell proliferation. c-Myc is constitutively and aberrantly expressed in more than 70% of human cancers, with many of its target genes encoding proteins that initiate and maintain the transformed state [43].

**Rat sarcoma virus protein (RAS):** Ras proteins belong to the superfamily of small guanosine triphosphatases, or small GTPases, the small G proteins with intrinsic GTPase activity, that control various cellular pathways critical to cancer development. Ras (RAS) proteins are products of the most frequently mutated RAS oncogenes in human cancers. These proteins are frequently involved in transporting signals from cell surface receptors to different intracellular targets inside the cell and are very important targets in cancer therapy.

RAS can serve as transducer and bifurcation signaling proteins capable of changing the properties of the signaling process through multiple downstream pathways, including signaling pathways that reach the nucleus to stimulate gene expression for cell proliferation. It is often required in receptor tyrosine kinase (RTK)-activated signaling pathways involved in stimulating cell growth, proliferation, and differentiation. Mammalian cells express three different yet closely related Ras proteins, K-Ras, H-Ras, and N-Ras, whose mutational activation effectively promotes oncogenesis. The mutation frequency of different Ras isoforms in human cancers varies, and K-Ras is the most frequently mutated isoform leading to tumor formation, invasion, and metastasis in many cancers. The mutation rate for K-Ras is approximately 25% for all tumors but is found to mutate up to 80–90% in pancreatic ductal adenocarcinoma (PDAC). The treatment of PDAC, the most common form of pancreatic cancer and a leading cause of cancer-related death, has thus far been sparsely productive because of the TME, which possesses many stromal cells and a complicated extracellular matrix (ECM). Genomic analysis has recently revealed that PDAC harbors frequently mutated genes, including those encoding KRAS, TP53, CDKN2A, and SMAD4, which can strongly influence cellular processes and change the tumor microenvironment, which in turn affects cancer progression. Drug development to block K-Ras has been partially successful, similar to many other drugs, as affected cells develop resistance to inhibitors, a common problem encountered with drugs designed for cancer therapy. The study of K-Ras resistance mechanisms reveals that researchers may have to explore several different drug combinations to overcome resistance, and few such developments are in the pipeline. Researchers are tirelessly working to target K-Ras and other signaling intermediates associated with cancer to develop novel therapeutic agents for different cancers [44].

**Receptor tyrosine kinase (RTK):** A series of growth factors and their receptors are involved in cancer development and metastasis. RTKs are a class of cell surface receptors for many polypeptide growth factors, cytokines, and hormones that can play vital roles in cancer development. RTKs are receptors with kinase-like activities with specialized structural and biological features capable of dimerizing with other adjacent RTKs, leading to rapid phosphorylation of tyrosine residues on target molecules to initiate several downstream biochemical cascades in affected cells. Upon binding with their specific ligands or growth factors, RTKs, such as erythroblastic leukemia viral oncogene homolog (ErbB), the family of tyrosine kinase growth factor receptors, which includes ErbB1, also called epidermal growth factor receptor (EGFR) or human EGF receptor (HER 1), insulin-like growth factor receptor (IGFR), platelet-derived growth factor receptor (PDGFR), fibroblast growth factor receptor (FGFR), vascular endothelial growth factor receptor (VEGFR), or hepatocyte growth factor receptor (FGFR/MET), generally interact with src - homology 2 (SH2) containing target proteins to control crucial functions, such as cell growth, proliferation, differentiation, apoptosis, inflammation, and stress responses. These cellular processes can be critical for reciprocal interactions between tumors and stromal cells and play a central role in the control of tumor formation, angiogenesis, and metastasis [45]. The multifaceted role of RTKs makes them suitable candidates for selective targeting in cancer therapy, but their involvement in alternate pathway activation often presents serious challenges to anti-RTK therapy.

**Insulin-like growth factor receptor (IGFR):** IGFR is an RTK that binds to insulin-like growth factor (IGF) with high affinity and is an important factor in the growth, differentiation, and context dependent survival of healthy and diseased cells. The IGFR pathway ligands, IGF-1 and IGF-2, and their receptors, primarily IGF-1R play important roles in the anchorage-independent growth of cells, which may enable cancer cells to survive and grow in the absence of anchorage to the ECM and neighboring cells. High gene expression levels of IGF-1 and IGF-1R are associated with the upregulation of pathways supporting cell growth and survival, cell cycle progression, angiogenesis, and metastatic activities during cancer development and are considered essential in many cancer types [46].

**Hepatocyte growth factor receptor (HGFR/MET):** HGFR/MET is a cell surface receptor tyrosine kinase encoded by the mesenchymal–epithelial transition factor (MET)gene in different cell types, including epithelial and endothelial cells, and plays an essential role in tumor growth and metastasis. Hepatocyte growth factor (HGF), which was first discovered for its growth-stimulating activity on cultured hepatocytes, is a multipotent cytokine that is primarily secreted by resident fibroblasts in the organ microenvironment and is associated with certain molecular networks and pathways within cells that play essential roles in maintaining homeostatic process in the body. HGF is the only ligand for the cell surface receptor, but it can relay messages through several key downstream pathways that can stimulate cell proliferation, differentiation, survival and migration and is commonly associated with processes such as organogenesis, chondrogenesis, hematopoiesis, blood vessel formation and wound healing. HGF activity is low in normal cells, but alterations in HGF and/or MET expression patterns have been implicated in the growth of both hematologic and solid cancers. Cancer-associated fibroblasts (CAFs) are activated fibroblasts that are key constituents of the tumor stroma and HGF is a major component of their secretome. The increased activity of HGF leads to abnormal cell proliferation and dissemination of transformed cells from the origin of tumor formation through angiogenesis and, these cells metastasize to distant parts of the body. The HGF/MET axis is a potential means to mitigate cancer progression and therapeutic strategies are evolving to effectively target this pathway in cancer treatment [47].

**G-protein-linked receptors (GPCRs):** GPCRs are serpentine transmembrane proteins that form the largest group of cell surface receptors and are linked to heterotrimeric GTP-binding proteins (G proteins) to mediate responses to various extracellular signaling molecules, including hormones, neurotransmitters and local mediators such as cytokines, chemokines and growth factors. All members of the GPCR superfamily have similar structures, containing an extracellular amino terminus, seven transmembrane α-helical domains, and an intracellular carboxy terminus, and the same signaling molecule or ligand can activate many different receptors making them the most likely targets for drug therapy. There are approximately 1,000 different GPCRs associated with humans, and each one is highly specific to a particular ligand. G protein-mediated networking and signaling are the most important features of GPCRs, which are initiated by ligand-GPCR interactions on the cell surface and play crucial roles in different physiological processes. There are different types of G proteins that specifically associate with a particular set of membrane receptors to mediate responses to signaling molecules, and can also interact with molecules in other cellular pathways, such as RTKs and ion channels to expand the landscape of their functions and minor defects in associated pathways due to changes in ligand concentration and/or alterations in receptor protein expression can lead to many pathophysiological conditions, including cancer.Trimeric G proteins remain attached to the cytoplasmic face of the plasma membrane and serve as message relay centers in the cell to transmit signals by coupling receptors to different enzymes or ion channels in the cell membrane. An activated receptor causes the trimeric G protein to dissociate, stimulating its components in different ways, and the GTP-binding protein subunit functions as a switch that can be turned on or off by ligand-receptor interactions on the cell surface, which is crucial for GPCR-mediated signaling. Activated G proteins target various enzymes that produce second messengers, such as cyclic AMP (cAMP), diacylglycerol (DAG), and inositol 1, 4, 5-triphosphate (IP3), as well as ion channels that transport certain ions to serve as second messengers. The generation of second messengers influences various intracellular processes, including the activation of protein kinases such as cyclic AMP-dependent protein kinase, also known as protein kinase A (PKA), and protein kinase C (PKC), which can phosphorylate many regulatory proteins leading to the timely regulation of target genes.Numerous studies have revealed that G proteins activated by GPCRs control many aspects of cancer progression, including tumor growth, cell survival, invasion, migration, and metastasis. Notably, approximately half of all known drugs actively target GPCRs, as they correspond to more than 30% of all identified drug targets, and genomic studies continue to reveal an increasing number of new family members whose detailed studies could lead to the identification of many potential drug targets for cancer treatment [48].
Steroid hormone receptor (SHR): SHRs are intracellular transcription factors that control endocrine modulatory mechanisms and play essential roles in normal cell growth, differentiation, and maintenance of tissue homeostasis. Several pathological conditions have also been associated with SHRs, including cancer. Many cancers, such as breast cancer, prostate cancer, ovarian cancer, endometrial cancer, lung cancer, leukemia and lymphoma, can be stimulated by steroid hormones via SHRs to grow and metastasize and are called hormone sensitive or hormone receptor positive cancers. SHRs belong to the nuclear receptor superfamily of transcriptional regulators and respond to specific steroids/hormones via paracrine or endocrine mechanisms at the level of gene regulation. As nuclear receptors, SHRs are modular proteins with three major functional domains, the N-terminal transactivation domain, a central DNA binding domain (DBD), and the C-terminal ligand binding domain (LBD). Several crucial intracellular pathways have been revealed for the actions of SHRs which primarily require receptor activation via ligand binding, followed by interaction with coactivators and the target gene for realization of the effect, thus they are important therapeutic targets in cancer treatment [49].

**Nuclear factor erythroid 2 (NFE2)-related factor (Nrf2):** Nrf2 belongs to the CNC (cap‘n’collar) family of proteins, a group of basic leucine zipper (bZip) transcription factors encoded by basic leucine zipper (bZIP) genes, which serve as master regulators of the cellular antioxidant response. Recent studies have revealed many new roles for Nrf2 in the regulation of essential cellular processes through interactions with other pathways within cells, thus establishing it as a truly pleiotropic transcription factor involved in carcinogenesis. Originally recognized as a target of chemopreventive agents to help prevent cancer, its protective role was altered in 6–7% of cancer cases. A growing body of evidence has revealed that the Nrf2 pathway is involved in the deregulation of the cellular metabolism, apoptosis, and self - renewal capacity of cancer stem cells and is thus an important driver of cancer progression, metastasis, and drug resistance [50].

**B-cell lymphoma-2 (Bcl-2):** The Bcl-2 oncoprotein is primarily a cell death regulatory protein that controls whether a cell lives or dies via apoptosis. It is a member of a family of regulatory proteins that are actively involved in the regulation of cell death via all major pathways, including apoptosis, autophagy, and necrosis, and serves at the critical junction of multiple pathways with crucial roles in oncogenesis. Aberrant expression of the BCL2 gene may prevent the death of cancer cells and is frequently implicated in prolonged cell survival and therapy resistance in human cancer. The Bcl-2 family of proteins forms subgroups, one of which may inhibit cell death and prolong cell survival by limiting apoptosis, whereas other induces cell death by inducing apoptosis or autophagy. The gene encoding the Bcl-2 protein is located on chromosome 18 but can be transferred to different chromosomes, as can be observed in many cancer types. Increased expression of pro survival proteins or an abnormal reduction in death-inducing regulatory proteins, resulting in strong inhibition of apoptosis and other related catabolic activities, is frequently observed in many cancers.

Resistance to apoptosis is a key development in several hematological malignancies and is attributed to the upregulation of pro survival Bcl-2 proteins. The important role of Bcl-2 family proteins in cancer development renders them potential targets for the treatment of different cancers, including Hematologic cancers and solid tumors. Changes in Bcl-2 activity with concurrent changes in other crucial proteins, such as c-Myc or p53, appear to be strongly associated with cancer progression. Recently developed inhibitors of pro survival Bcl-2 proteins, termed BH3-mimetic drugs wherever applicable, have been used as novel agents for cancer treatment [51].

## 6. Signaling pathway deregulation and prospective targets in Cancer therapeutics

Cancer growth and progression are dependent on complex interactions between tumor cells, surrounding stromal cells and the ECM present in the TME. However, the root cause underlying cancer progression remains genetic and epigenetic alterations linked to the regulation of cell growth and proliferation, cell adhesion, immune suppression, cell death, differentiation, and overall genomic stability of the affected cells, leading them to grow and proliferate uncontrollably beyond barriers [52]. It is ultimately driven by dysregulated molecular mechanisms involving tumor suppressor genes, oncogenes, growth factors, cell adhesion molecules, and molecules of the immune system, such as cytokines and chemokines, that may vary among different cancer types and stages. The cell signaling network, as the foremost system of communication between cells and their surroundings that involves a variety of chemical and mechanical signals and networks of intracellular proteins to constitute different molecular signaling pathways, is worth considering here, as all the essentials of cellular behaviors, such as cell growth and proliferation, cell polarity, cell metabolism, differentiation, survival, and migration, can be guided by the components of these pathways working in a collaborative manner inside the cell [53]. A signaling pathway, in general, constitutes a cascade or chain of proteins that communicate signals from extracellular signaling molecules or other external stimuli, through the receptor on the cell surface to target genes in the nucleus of the cell and results in the expression of certain proteins that produce some changes in cell behavior, such as cell division and differentiation. Together, different signaling pathways maintain internal circuitry inside cells guided by external stimuli such as growth factors and cytokines, enabling them to sense whether their state of attachment to the ECM and other cells is appropriate, and if different growth factors, hormones, and cytokines guide them to proliferate or differentiate, they can move, stay put for now, or commit to cell death by apoptosis or autophagy [54]. Almost all gene modifications can be related to one or more of these signaling pathways that are deregulated in the affected cells to acquire hallmark properties of cancer. Cancer cell signaling typically involves altered expression of the components of the signaling network, which include many secreted protein receptors, growth factors and cytokines, protein kinases, phosphatases, different cytoplasmic proteins, and transcription factors, leading individual cells to respond to genomic changes with appropriate physiological behaviors.

Cell division is regulated mainly by a group of extracellular growth factors that signal that resting cells divide by exploiting their intrinsic regulatory processes. Cytokines signal immune cells to mount coordinated attacks on invading bacteria and viruses and play essential roles in cancer prevention. Thus, signals propagated by growth factors and cytokines can simply tell individual cells to divide or not under particular conditions whose alterations could lead to the pathophysiology of cancer.

The earliest information regarding the relationship between cancer and growth factors came from the observation that normal cells in culture often require serum for proliferation, whereas cancer cells have a much lower requirement for serum. Serum is known for providing growth factors, among other ingredients needed for the overall regulation of the cell cycle. The other indication revealed that gene mutations found in cancer cells cause changes in cell behaviors very similar to those related to the activities of growth factors and their receptors. Oncogenic mutations disrupt the cellular circuits that control cell adhesion and signaling, enabling cells that carry them to overproliferate and invade other tissues in an uncontrolled fashion. Many of these mutations have been directly linked to growth factors and their receptor proteins, which are involved in tumor growth, angiogenesis, invasion, and metastasis [55,56]. Importantly, one type of cell membrane receptor can mediate many different downstream intracellular pathways, and one pathway can also be activated by several upstream surface receptors, revealing common signaling components in multiple signaling pathways. For example, RTKs, such as EGFR, IGFR, PDGFR, FGFR, VEGFR, HGFR, or GPCR, can activate the mitogen-activated protein kinase (MAPK) cascade, whereas widely studied RTKs, such as the EGFR/HER family of receptors, can initiate different signaling pathways, including the MAPK, phosphoinositide-3-kinase (PI3K), and mammalian target of rapamycin (mTOR) pathways, which are commonly involved in the regulation of cell growth, proliferation, differentiation, and survival. This feature of the signaling process evidently presents the option for crosstalk between components of different signaling pathways at different stages of the cellular process. A molecule participating in crosstalk can affect the activation of alternate signaling pathways, and receptors can also have an altered ability to bind to ligands, which can swiftly lead to cancer manifestation. As generally observed, most cell signaling pathways contribute to the development of cancer, and very few cancer types arise from the deregulation of a single pathway. Breast cancer can arise from elevated expression of the estrogen receptor (ER), EGFR/HER, or IGFR, but in many cases, molecules and intermediates of multiple signaling pathways can be interactively involved in this process. In this way, many signaling molecules affecting cancer cells together could be considered to create elaborate integrated circuits within the cell, derived from the usual signaling circuits that operate in normal cells. The transformed intracellular circuit can be divided into distinct subcircuits specializing in specific cellular activities to promote hallmark features of cancer (Fig. 1) [57].

Signal transduction pathways that lead to tumor growth, cancer cell migration, metastasis, and drug resistance are often complex processes, as cancer cells typically develop abnormalities in multiple signaling pathways or rely on crosstalk between different pathways and some redundant pathways for the maintenance of growth and survival. As cancer progression involves alterations in signaling pathways due to mutations in relevant genes, it is worth considering that therapeutic intervention that takes into account the biology of affected cells can pave the way for very effective cancer treatment [58,59]. Importantly, in clinical practice, targeting a single intermediate or pathway results in considerable recovery, possibly because it impedes the synergistic signaling process of disease progression. Nevertheless, the constitutive activation of a molecular event that contributes to cancer development can be sustained by different mechanisms, and strategies to inhibit multiple targets or redundant pathways simultaneously with molecular-targeted agents could prove to be an even more effective way to treat cancer and overcome resistance in cancer therapy [60]. This approach has indeed been used with anticipated outcomes in some forms of cancer, indicating the need for more research in that direction.

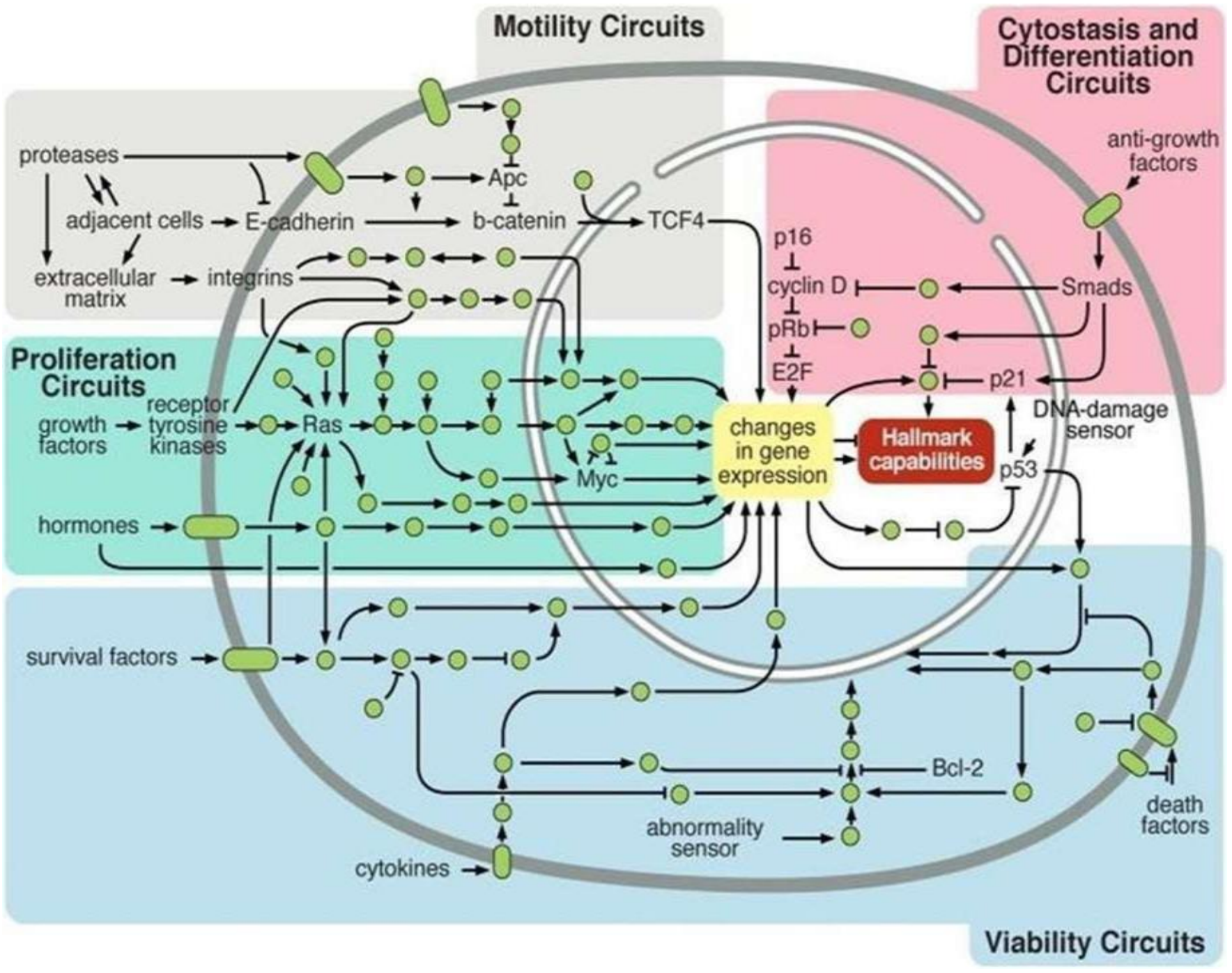

**Figure 1.** Intracellular Signaling Networks Regulate the Operations of the Cancer Cell.
An elaborate integrated circuit operates within normal cells and is reprogrammed to regulate hallmark capabilities within cancer cells. Separate subcircuits, depicted here in differently colored fields, are specialized to orchestrate various capabilities. At one level, this depiction is simplistic, as there is considerable crosstalk between such subcircuits. In addition, because each cancer cell is exposed to a complex mixture of signals from its microenvironment, each of these subcircuits is connected with signals originating from other cells in the tumor microenvironment. (Hanahan and Wienberg [57]. With permission from Elsevier)

The challenge of identifying the genes and signaling molecules relevant to different cancer types by cutting-edge technologies remains an essential part of cancer research and is most likely to help vulnerable people receive precisely designed treatments for cancer.
The representative signaling pathways involved in cancer cell reprogramming and the scope for therapeutic targeting of signaling molecules and intermediates for efficient cancer treatment are briefly discussed here.

**Ras/Raf/MAPK signaling pathway:** The mitogen-activated protein kinase (MAPK) signaling cascade is the evolutionarily conserved signaling pathway, which is the main route by which extracellular growth factors transmit signals to cells that regulate a wide variety of cellular processes, including cell proliferation, differentiation, apoptosis, and stress response, and abnormalities in this pathway are common in many cancer types [61]. This cascade is the key downstream effector of Ras GTPase which involves rapidly accelerated fibrosarcoma (Raf) kinases, MAPK/ERK protein kinases (MEKs), and mitogen-activated protein kinases (MAPKs), also called extracellular signal-regulated kinases (ERKs), The binding of extracellular growth factors such as EGF or FGF to appropriate cell surface receptors stimulates Ras GTPase activities, whichin turn activate Raf kinases. The RAF kinase phosphorylates and activates MEKs, resulting in the activation of ERKs. Activated ERK relays signals downstream of transcription factors or other gene regulatory proteins, resulting in the expression of target genes, which has been the subject of intense scrutiny in the treatment of cancer. Most growth factor receptors, such as the TGF-β receptors, EGFR, IGFR, PDGFR, FGFR, VEGFR, and HGFR, can activate Ras, ultimately leading to ERK activation. Importantly, Ras GTPase may act as a molecular switch that controls the activation and regulation of related cellular pathways responsible for different cell behaviors critical to cancer development [62]. Furthermore, the mutational activation of Raf in human cancers supports the important role of this pathway in oncogenesis. Studies with selected inhibitors against targets in this cascade have shown positive results, such as growth inhibition, antiangiogenic effects, and suppression of metastasis in cancer cell lines and animal models. These results reveal that this strategy is effective at inhibiting cancer cell proliferation and survival, and more clinical trials and validations are ongoing for the efficacious treatment of cancer [63].

**PI3K/Akt/mTOR signaling pathway:** It is a highly conserved signaling pathway in eukaryotes that plays an important role in physiological and pathological development. A variety of cellular processes are stimulated by this pathway, ultimately leading to increased cell growth, proliferation, and loss of apoptosis which promote overall cell proliferation and survival, earning it the nickname of the 'survival pathway'". This pathway can be activated by a variety of factors, such as cytokine receptors, GPCRs, RTKs, and integrins, and regulates several cellular and metabolic activities that lead to cell growth and survival.
Phosphatidylinositol (PI) is a unique membrane lipid that is phosphorylated by activated, PI3-kinase to generate phosphatidylinositol-3,4,5-triphosphate [PIP3], which works as the docking site for intracellular signaling proteins, bringing the proteins together into signaling complexes. The main PI3K effector protein kinase B (PKB), also known as Akt (Ak strain transforming, in relation to AKR mice), is an important signaling center with many critical target substrates, including mTOR, which is activated in the process of regulating different downstream targets to relay signals through the cell. The kinase protein mTOR is of particular interest because it works as a master regulator of cellular processes by participating in multiple signaling pathways inside the cell and is actively involved in cell growth, proliferation, autophagy, and apoptosis. The canonical pathway of mTOR activation depends on signaling through PI3K/Akt, although alternative non-Akt-dependent activation through the MAPK pathway is now well recognized. Activated mTOR can assemble into a variety of complexes to catalyze the phosphorylation of multiple targets, including Akt, protein kinase C (PKC), components of insulin-like growth factor receptor (IGF-IR) signaling, and the protein synthesis machinery to influence a variety of cell behaviors. Persistent mutational activation of the PI3K/Akt/mTOR pathway in the absence of different stimuli has been frequently observed in many cancers. Several mTOR inhibitors have also been developed to treat cancer, and some are being evaluated in clinical trials for approval [64,65]. In addition, phosphatase and tensin homolog (PTEN), a potent tumor suppressor, is a crucial component of mTOR mediated pathways that can also work independently as a phosphatase against phospholipids and proteins. Its primary target is PIP3, the direct product of PI3K, which is critically involved in the signaling process. Mutational deregulation of the PTEN/PI3K network has been associated with many cancer types, including familial cancers. It is a potential means of targeting PI3K-mediated signaling in cancer therapeutics [66]. Adaptive resistance to pathway inhibitors is common, and combination therapy, if well tolerated, may produce favorable anticancer results [67].

**JAK/STAT signaling pathway:** The Janus kinase (JAK)/signal transducer and activator of transcription (STAT) signaling pathway is actively involved in the regulation of essential cellular activities, such as proliferation, survival, invasion, inflammation, and immune deregulation, which are associated with cancer progression and metastasis. There are seven different signal transducers and activators of transcription (STAT) family proteins in mammals: STAT 1, 2, 3, 4, 5A, 5B, and STAT 6. The Janus kinase (JAK) family comprises four different members: JAK1, 2, 3, and Tyk (tyrosine kinase). This pathway largely involves cytokine signaling, which is closely related to the activities of T and B cells and is often linked to the development of hematological malignancies. When a cell is exposed to cytokines such as interleukin-6 (IL-6) or interferon-gamma (IFN-g), JAK kinases associated with cytokine receptors are activated to phosphorylate and activate STATs. STAT family members, especially STAT3 and STAT5, are involved in cancer progression, whereas STAT1 plays the opposite role by suppressing tumor growth. The target genes of STAT5 may regulate processes such as cell cycle progression, survival, and self-renewal by binding to growth factors and cytokines, and constitutive activation of the pathway leads to high-level expression of genes and proteins, resulting in different forms of cancer [68,69]. It can ultimately be mediated through the suppression of p53 activity, crosstalk with NF-kB signaling, or expression of Runt-related transcription factor (RUNX) family proteins, leading to inflammation and cancer [70]. The activation of the JAK/STAT pathway can be controlled by suppressors of cytokine signaling (SOCS) family proteins, whereas other inhibitory proteins and phosphatases may also contribute to inhibiting the activated state. The upregulation of JAK/STAT proteins, as well as the reduction in different SOCS proteins, are associated with different malignancies, including solid tumors. This signaling pathway has also been associated with the development of tumor tolerance, as hyperactivation of the pathway often leads to an increase in gene expression, resulting in increased activity of regulatory T cells (Tregs), a specialized subpopulation of T cells that work to limit T-cell proliferation and cytokine production, thereby resulting in the suppression of the immune response and maintenance of self-tolerance. These specificities of the signaling pathway provide options for effective drug development against pathway intermediates with fewer side effects. Many JAK and STAT inhibitors have been tested for their efficacy in cancer treatment, and a few of these inhibitors have been shown to be clinically relevant. Efficiently targeting the JAK/STAT signaling pathway remains an intriguing strategy in cancer therapy [71,72].

**TGF-β/SMAD signaling pathway:** This pathway involves transforming growth factor-beta (TGF-β) superfamily proteins that serve as multifunctional secreted cytokines whose activities can be deregulated in many diseases, including cancer. TGF-β signaling is known to control many different biological processes, including cell proliferation, differentiation, migration, and apoptosis, and plays context-dependent roles in carcinogenesis. SMAD proteins are the main signal transducers for the canonical pathway of TGF-β signaling. It comprises a family of structurally similar and well-conserved transcription factors that can relay extracellular signals directly to the nucleus and are critically important for regulating cell development and growth. TGF-β initially functions as a tumor suppressor through the SMAD-mediated pathway when TGF-β/SMAD-dependent p15/p21 induction or c-MYC suppression works well to maintain growth arrest, cell differentiation, and apoptosis.

However, the situation could be the opposite if SMAD-dependent suppression becomes ineffective under the influence of certain oncogenic mutations or other signaling pathways, and the role of TGF-β could become antiapoptotic, EMT inducers, and carcinogenic. SMAD inactivation under such circumstances convincingly explains the situation-based role of TGF-β in different malignancies. Furthermore, the classical SMAD-independent pathway of TGF-β receptors may engage in crosstalk with other signaling pathways, such as the Wnt/β-catenin, Ras/RAF/MAPK, and PI3K/Akt/mTOR pathways, to play vital roles in carcinogenesis, and a proper understanding of the TGF-β signaling pathway in cancer progression would resolve controversies related to these signaling pathways [73,74]. The wide range of functions associated with TGF-β during cancer progression is now clear, which has led to the development of multiple therapeutic agents targeting different intermediates of the signaling pathway, and a combination of drugs may produce even better results against TGF-β -mediated recurrent and metastatic cancer [75,76].

**The Hippo signaling pathway:** This pathway is an evolutionarily conserved major signaling pathway that was originally identified in fruit flies (Drosophila melanogaster) and controls contact inhibition and organ size development. It is a serine/threonine kinase signaling cascade, and its dysregulation has been implicated in many cancer types. Contact inhibition enables normal cells to cease growth and proliferation when in contact with each other, and an absence of this property can lead affected cells to proliferate uncontrollably, resulting in malignant growth. The canonical Hippo pathway comprises a kinase cascade and related regulators that work together as a repressive system involving phosphorylation and inhibition of the two transcription coactivators YAP and TAZ as downstream effectors to execute their role in the regulation of organ size and tissue homeostasis. Phosphatase and protein ubiquitination modulate the activities of the coactivators in the cascade and can also be regulated by the cytoskeleton for their role in the signaling process. When dephosphorylated, YAP/TAZ translocates into the nucleus and interacts with other transcription factors to induce gene expression, leading to cell proliferation and inhibition of apoptosis. The regulation of YAP1/TAZ may be influenced by many other molecular events, including crosstalk with Wnt/β-catenin signaling, and is mostly oncogenic. The core activity of this pathway is controlled by cell density, polarity, and energy requirements as well as by ECM stiffness and shear stress, which together can regulate contact inhibition and related development; thus, its activities can be regulated at multiple levels and widely implicated in angiogenesis and chemoresistance [77]. Cell proliferation and stem cell self-renewal can be directly attributed to contact inhibition governed by this signaling pathway. The noncanonical Hippo pathway operates in tight and adherens junction complexes to control their localization and activity within the cell. Several studies suggest that overexpression of the components of the Hippo pathway contributes to aberrant cell cycle regulation, leading to cancer development. The exact role of the Hippo pathway in cell cycle regulation is not fully understood, but an in-depth exploration of this process could provide effective therapeutic options for cancer treatment. The properties of the extracellular signaling and membrane receptors involved with the pathway remain to be fully known, yet drugs targeting the components of this pathway are under investigation for their efficacy in cancer therapy [78.79].

**Wnt/β-catenin signaling pathway:** This pathway is one of the key signaling cascades involved in the regulation of cell growth and cell polarity during development. It is typically associated with stemness and can be frequently implicated in carcinogenesis. The signaling pathway begins with Wnt ligand–protein binding to the extracellular domain of a Frizzled (Fz) family receptor, a distinct family of GPCRs that generally do not involve the activation of G proteins, to relay signals through the cell via different paths to influence a variety of cellular mechanisms critical to cancer development. The Wnt pathway has been formally divided into the β-catenin-dependent canonical pathway and the β-catenin-independent, noncanonical planar cell polarity (PCP) signaling pathway and the Wnt/calcium pathway.

Canonical Wnt signaling is a genetic pathway that promotes normal cell growth and requires meticulous control of a tumor suppressor gene called adenomatous polyposis coli (APC), which functions to limit the activation of β-catenin, preventing excessive cell growth and tumor formation. The APC/β-catenin pathway is a highly regulated process that involves many different proteins. APC itself is a negative regulator and a Wnt antagonist that binds to a variety of proteins, including β-catenin. It is an essential component of the cytoplasmic protein complex that targets β-catenin for proteasomal destruction. Furthermore, MYC and cyclins are important transcriptional targets of this pathway, indicating that they overlap with several tumor-promoting pathways. Mutations that prevent the degradation of β-catenin, including certain mutations in β-catenin or the APC component of the β-catenin destruction complex, distort the regenerative pathway to contribute to cancer progression and metastasis [80]. Deregulation of the signaling pathway results in alterations in cell growth and survival, maintenance of cancer stem cells, metastasis, and immune control, which have been linked to both solid and hematological tumors. The activation of the noncanonical pathway generally involves the recruitment of Rho family small GTPases, which leads to enzymatic rearrangements of the cytoskeleton and/or certain transcriptional activation of effector proteins. Both of these pathways essentially require the binding of Wnt proteins to Frizzled receptors to execute their functions. Wnt/Ca2+ signaling is followed by G protein-activated phospholipase C activity, leading to intracellular calcium flux and downstream calcium-dependent cytoskeletal rearrangement and/or transcriptional responses. The Wnt signaling pathway is a crucial mediator in maintaining tissue homeostasis, stem cell populations for tissue repair, and wound healing and is frequently involved in the manifestation of many cancer types. Mutations in the APC gene are observed in approximately 80% of colon cancers where cancer stem cells (CSCs) are thought to play a critical role in metastasis and relapse, indicating the role of this signaling pathway in maintaining CSCs. The role of Wnt signaling in cancer immune evasion and drug resistance is well recognized, and identifying tumor-specific signaling intermediates as targets for drug action can be crucial for effective cancer therapy. Many different agents effectively targeting molecules of this signaling pathway are being explored for the efficacious treatment of different cancer types [81,82].

**Hedgehog (Hh) signaling pathway:** The Hh pathway is an evolutionarily conserved signaling pathway and one of a few signaling pathways frequently involved in intercellular communication. It is a key regulator of embryonic development that controls cell patterning, proliferation, and differentiation for organ development in mammals as well as in the regeneration and maintenance of tissue homeostasis. This pathway has frequently been associated with birth defects, stem cell renewal, and cancer. Hh signaling depends on three transmembrane receptor proteins. Specifically, Patched, iHog, and Smoothened. Hh proteins are encoded by at least three genes in vertebrates: Sonic, Desert, and Indian hedgehog. Hh functions through a signaling cascade in a context-dependent manner to regulate the balance between activator and repressor forms of glioma-associated oncogene (Gli) transcription factors.

There are three different forms of the transcription factor Gli1. Gli2 and Gli3 are present in vertebrates and may undergo proteasomal processing similar to that of the Wnt pathway to exert their effects in response to appropriate signals. The activated form of Gli moves to the nucleus to bind to its promoter, leading to the transcription of target genes. Mutational changes that lead to excessive activation of the Hh pathway have been implicated in different malignancies. Communication between Hh and major signaling pathways, such as the Wnt, Notch, and TGF-β pathways, plays crucial roles in the pathophysiology of cancer. Several Hh signaling pathway inhibitors have been developed for a range of cancers, and a few agents are thought to be highly effective for patients with recurrent and advanced cancers [83].

The Notch signaling pathway: It is a contact-dependent signaling pathway that plays a major role in controlling cell fate decisions and regulating pattern formation during the renewal and development of most tissues and performs major tasks during the embryonic development of animals. Signaling is mediated through the Notch receptor protein, a single- pass transmembrane protein that undergoes successive proteolytic cleavage steps upon activation to perform its action. Notch is activated in a contact-dependent manner by a specific signal protein called Delta, which is present on neighboring cells and leads to the cleavage and release of its cytoplasmic tail, the notch intracellular domain (NCID), which moves to the nucleus, where it regulates the expression of target genes [84]. Notch signaling is associated with the regulation of many cellular processes, such as cell proliferation, survival, differentiation, and apoptosis, through cell-to-cell communication crucial to the development of many tissues. The signaling pathway is a key regulator of self- renewal and differentiation in many cell types and is known to be an important regulator of hematological processes. Notch acts as a context-dependent binary cell fate-determining pathway, and its hyperactivation has been implicated in the oncogenic stimulation of many solid and hematological cancers.

The Hh and Notch signaling pathways are active regulators of communication between cells and are actively involved in the regulation of EMT, which is critical for organ development, regeneration, stem cell maintenance, and tissue homeostasis. The self- renewal potential of cancer stem cells (CSCs) is attributed to these signaling pathways, which are crucial for maintaining CSCs in the tumor mass and cause disease progression, recurrence, and chemoresistance. Importantly, the Hippo pathway has been found to repress Wnt signaling, which can induce cancer stem cell activities. In addition, alterations in Wnt signaling are known to influence the Hg and Notch pathways alternatively, which can be intrinsically related to the maintenance of CSC properties [85]. Thus, the components of one signaling pathway could influence the performance of the other pathways to synergistically maintain the activities of the CSCs involved in cancer development. This observation presents the option to identify signaling intermediates with confirmed hyperactivities as potential targets in anti-CSC drug discovery for effective cancer treatment. Selective targeting of these pathways, along with other proliferative pathways, such as the PI3K/Akt or RAS/RAF/MAPK pathways, could prove to be an effective strategy for combination therapy of cancer [86, 87].

**The NF-κB signaling pathway:** This pathway involves nuclear factor kB (NF-kB), a family of transcription factors that control the expression of specific genes to regulate multiple physiological activities. Its main role is to mediate immune and inflammatory responses, in addition to being involved in various cellular activities such as cell adhesion, growth and differentiation, proliferation, autophagy, and inhibition of apoptosis. It therefore plays a critical role in cancer progression by influencing cell growth, proliferation, survival, and immune responses.The IkB proteins bind to the resting NF-kB dimers, preventing it from binding to DNA. This pathway is initiated by the degradation of IkB proteins via IkB kinase (IKK) leading to NF-kB activation and, consequently, transcriptional activation. NEMO is a non-enzymatic scaffolding protein and a regulatory unit of the IKK complex, essential for NF- kB signaling. Importantly, this signaling can be mediated via both the NEMO-dependent canonical pathway and the NEMO-independent noncanonical pathway. The canonical pathway is thought to be involved in immune responses and immunosurveillance, whereas the noncanonical pathway is associated with immune cell differentiation and maturation, and secondary developmental activities. The canonical and non-canonical pathways are generally distinct, but studies reveal numerous cross-talk mechanisms that link them, such that both pathways could result in a single NF-κB system [88]. This cross-talk could involve regulatory control of NF-κB monomer expression and interdependent processing of regulatory proteins. Constitutively activated NF-κB signaling may lead to inflammation- related disorders, and its role in pathological inflammation and cancer development is well recognized [89]. Furthermore, NF-κB signaling is associated with epithelial–mesenchymal transition (EMT), which frequently occurs during tumor progression and metastasis. E- cadherin is a well-known tumor suppressor protein; the regulation of the adhesive activity of E-cadherin present at the cell surface is important in cancer, and its repression by NF-κB is attributed to EMT induction. NF-kB has also been implicated in EMT and metastasis through the activation of EMT master-switch transcription factors and is highly invasive. Evidence suggests that the reversal of EMT is triggered by the inhibition of NF-kB signaling. Moreover, activated NF-κB pathway may contribute to anti-apoptotic activation and ECM degradation in addition to E-cadherin-mediated EMT, to contribute to tumor growth, invasion, and metastasis [90]. NF-κB signaling molecules also communicate with many other signaling pathways, as crosstalk can be mediated by intermediates, such as STAT3 and, GSK3-β, p53, p38, PI3K, or proinflammatory TGF-β proteins, which modulate NF-κB transcriptional activity. Thus, targeting the NF-κB signaling pathway represents an attractive approach to anti-inflammatory and anticancer therapies, and inhibitors have been developed to block different steps of NF-κB signaling for cancer treatment [91].

**The Aurora kinase signaling pathway:** This signaling involves Aurora kinases which are a family of highly conserved serine/threonine kinases that are essential for cell division and are active in rapidly dividing cells such as embryonic cells, hematopoietic cells, germ cells or healing tissues, while their activities remain very low or absent in normal adult tissues due to the near absence of cell growth and proliferation there. Cell division which requires equal division of chromosomes and cytoplasm between daughter cells is strictly regulated by a series of serine/threonine kinases, mainly Cyclin-dependent kinases (Cdks), Polo like kinases (Plks) and Aurora kinases among which Aurora kinases are the most important and indispensable as they can regulate Plks and Cdks at different stages of cell cycle progression to accomplish the end results. First discovered nearly thirty years ago in Drosophila mutants and named for their unusual spindle morphology such as that of the northern lights (aurora borealis), Aurora kinases have been widely studied for their roles in cancer progression and therapy owing to their importance in cell cycle regulation [92]. This family includes the Aurora A (AURKA), Aurora B (AURKB) and Aurora C (AURKC), encoded by the aurora a, aurora b and aurora c genes, respectively, and have common structural features containing an N-terminal domain, a protein kinase domain and a C-terminal domain. The kinase domain shares a high degree of sequence identity and must be activated by phosphorylation by certain protein cofactors, target proteins, or autophosphorylation by clustering kinase molecules for their enzymatic activities. Aurora kinases exhibit chromosomal localization and are crucial for the regulation of cell cycle entry into the M phase and appropriate chromosomal segregation during cell division. These kinases are expressed in a timely manner during cell cycle progression and undergo proteasomal degradation later in the process. AURKA can initially be found at the centrosomes where it is involved in centrosome maturation and separation, and shifts to spindle microtubules for spindle assembly for the completion of cell division. The level of AURKA increases throughout the S and G2 phases and peaks during the M phase to regulate mitotic entry and cell cycle progression. AURKB is located at the kinetochore and is typically associated with chromosomal alignment and segregation during the M phase and is necessary for cytokinesis and exit from mitosis. AURKC is not expressed in somatic cells but only in mammalian germ cells during meiotic cell division to regulate chromosomal segregation and cytokinesis. Abnormally expressed Aurora kinases have been frequently associated with abnormalities related to the cell division process, including genomic instability, chromosome number aberrations, and tumor formation. The lack of aurora kinase activity leads to failure of cell division whereas the overexpression of these kinases is associated with a number of cancers [93]. Studies have also revealed that aurora kinases can interact with members of other signaling pathways, such as the MAPK, PI3K/Akt, Wnt/β- catenin, or NF-κB pathways to promote cancer progression. The interaction of AURKA with proteins like p53 and MYC is closely linked to cancer incidence. All three Aurora kinases have been implicated in the development of both hematological and solid cancers and may promote carcinogenesis by inducing cell proliferation, cancer stem cell formation, EMT, and/or downregulation of apoptosis. Anticancer agents that selectively target Aurora kinases and their associates including Plks and Cdks have potential anticancer effects. Many specific Aurora kinase inhibitors (AKIs) are being tested and found to be highly effective and low in toxicity in the treatment of cancer [94].

**The cGAS–STING pathway:** The cyclic GMP–AMP synthase (cGAS)-stimulator of interferon genes (STING) signaling pathway represents a key cellular process that controls inflammatory responses in the presence of foreign particles on the basis of dsDNA recognition through pattern recognition receptors (PPRs) and thus regulates overall preparedness for the cell to withstand adversity caused by infection or injury. The binding of cGAS to double-stranded DNA (dsDNA) induces the catalytic activity of the synthase and leads to the production of 2′ 3′ cyclic GMP – AMP (cGAMP), a second messenger molecule that quickly binds to stimulator of interferon genes (STING) dimers localized at the endoplasmic reticulum (ER) membrane, which are then released to undergo further processing, finally resulting in the expression of type I interferons, interferon-stimulated genes (ISGs), pro-apoptotic genes and several other pro-inflammatory cytokines and chemokines [95]. STING also binds and stimulates IKK, triggering the transcriptional activation of NF-κB, which promotes noncanonical NF-κB responses. This signaling outcome limits type II interferons and the canonical NF-κB pathway as critical, negative regulators of STING effector mechanisms, which can have important biological consequences related to cancer immune evasion and metastasis. cGAS–STING signaling may also induce autophagy and additionally communicate via p53, MAPK p38, and STAT3 signaling in a context-dependent manner [96]. These findings reveal the complex role of this signaling pathway in the regulation of cell behaviors, and mutations associated with this pathway have often been implicated in cancer progression. This signaling pathway is important in cancer immunotherapy, and inhibitors of the related pathways are being used for targeted cancer therapy [97].

**Rho/ROCK signaling pathway:** Components of the Rho/Rho-kinase (ROCK) signaling pathway are potential regulators of the actin cytoskeleton and its dynamics inside the cell. ROCKs (ROCK1 and ROCK2) belong to the AGC (PKA/PKG/PKC) family of serine/threonine-specific protein kinases, which are downstream effectors of the small guanosine triphosphatases (GTPases) RhoA, B, and C and actively participate in a variety of cellular activities controlled by the actin cytoskeleton, including cell polarity, cell contraction, cell cycle progression, proliferation, motility, and invasion. Aberrant Rho/ROCK signaling has been implicated in several cancer types owing to its ability to increase tumor growth, cell migration, metastasis, and extracellular matrix remodeling [98]. Molecular inhibitors with high clinical value for the treatment of advanced solid cancers are being developed to target ROCK1, ROCK2, or both. Moreover, the different activities of ROCK in the immune system make it a potential target in cancer immunotherapy, so ROCK is thought to be of great value in cancer therapeutics. A deeper understanding of this pathway may add new dimensions to future precision cancer therapy [99].

## 7. High-throughput technologies and computational strategies for targeted cancer therapy

High-throughput sequencing technologies, also known as next-generation sequencing (NGS), are comprehensive terms used to describe technologies that sequence DNA and RNA rapidly and cost-effectively. It has revolutionized the fields of genetics and molecular biology and aided in the study of biological sciences as never before [100]. NGS can swiftly reveal the nature of genes and proteins thought to be associated with cancer, and the application of a few such evolving molecular techniques to the study of cancer has also provided cancer biomarkers over the years that have led to new advances in tumor diagnosis, prognosis, and treatment, which have proven to be immensely helpful in advancing precision oncology [101]. Cancer biomarkers simply refer to a variety of biomolecules, including transcription factors, cell surface receptors, metabolites, circulating tumor DNA/RNA, and secreted proteins produced by tissues as a result of cancer development. The identification of biomarkers is important from diagnostic, prognostic, and therapeutic perspectives because of increasing advances in our understanding of the cell and molecular biology of cancer development and possible improvements in therapeutic options for cancer treatment. As methods for studying cellular pathways have improved and the range of possible treatment options has expanded, cancer biomarkers are becoming important for accurately predicting how patients respond to specific treatment regimens, which is crucial for precision oncology. Circulating cell-free DNA (cfDNA), cell-free RNA (cfRNA), and extracellular vesicles or exosomes in the blood are abundantly released by cancer cells, which can be identified via liquid biopsy and are excellent sources of a variety of molecular markers. NGS-based molecular profiling of these markers can be used to gain crucial information regarding cancer development, including information on tumor heterogeneity. Definitive biomarkers can reveal disease prognosis, predict the likely response to specific treatments, the chances of recurrence, and survival, and can therefore play a critical role in the development of anticancer agents. Many diagnostic and prognostic biomarkers can also be used as potential therapeutic targets. There are many reliable prognostic, diagnostic and therapeutic markers recognized for cancer, and some are highly effective targets for cancer therapy [102].

Technologies using NGS have been developed that measure some characteristics of a whole family of cellular molecules, such as genes, proteins, or metabolites, and have been named by appending the term "-omics. Rapid progress in the development of NGS-based techniques for genomics, epigenomics, and transcriptomics has provided many valuable insights into the genetic mechanisms underlying cancer development. Multiomics refers to the approach where datasets of different omics groups are combined during sample analysis to allow scientists to read the more complex and transient molecular changes that underpin the course of disease progression and response to treatment and to select the right drug target for the desired results [103]. It forms the basis of precision medicine in general and is at the center of new developments in precision oncology research.

Tumor heterogeneity is a hallmark property of cancer development and broadly refers to the differences between tumors of the same type in different patients, the differences between a primary and a secondary tumor, and the differences in genomic and phenotypic profiles displayed by cells within a single tumor. Heterogeneity within a single tumor, referred to as genetic intratumoral heterogeneity (ITH), has been documented across most cancers as an outcome of genome instability and clonal evolution. Tumor heterogeneity appears to be a critical phenomenon in the history of individual cancers, as its translational importance may reflect tumor progression, disease recurrence, treatment response, and resistance. Recent investigations on drug resistance and tumor heterogeneity have confirmed the clonal organization of tumors as the underlying basis for drug resistance, thus indicating the need to fully understand the structure and dynamics of ITH to develop advanced treatment strategies for cancer [104]. Given that the cellular composition of a tumor is precisely known, the underlying mechanism of disease progression is understood, the molecules and pathways involved in the process are identified, a far more specific therapeutic strategy could be devised to achieve the desired result. This is the stated goal of precision oncology, and the emergence of single-cell technologies for biological analysis has become a crucial tool in this regard in recent years. This NGS-based techniques is equipped to provide a clear picture of tumor heterogeneity and reveal how structural changes in chromosomes can lead to the complex biological processes involved in cancer. Single-cell Technology can carry out single-cell measurements of a sample using single-cell multi-omics to study the complexity of gene function, disease development, and therapeutic response at single cell resolution for efficient cancer treatment. Single-cell multiomics now facilitates the simultaneous measurement of thousands of genes and proteins in thousands of ‘single’ cells from a single sample, allowing researchers to compare the genomes of individual cells to determine the mutational profile of the affected cells to better understand the molecular consequences of different variants present in the tumor [105,106].

As a matter of fact, the breakthroughs in high-throughput technologies in recent years have led to the rapid accumulation of large-scale omics cancer data and brought an evolving concept of “big data” in cancer analysis. It further highlights the need for advanced computational resources for analyzing complex data to bring new insights into critical problems of cancer treatment [107]. Artificial intelligence (AI) is a major advancement in computational strategy related to healthcare that encompasses many computational and bioinformatic technologies with the common goal of mimicking human intelligence to solve complex biological problems. It is based on the principle that human intelligence can be defined in a manner such that a machine can easily mimic and execute tasks from simpler to far more complex ones successfully [108,109]. Broadly referred to as computer programming, which is enabled to perform specific tasks, the term may be applied to any machine that displays traits associated with human intelligence, such as learning and problem solving. The combination of big data, bioinformatics, and artificial intelligence is thought to lead to notable advances in translational research in cancer [110,111].

The effective development of drugs for cancer treatment is a major problem in cancer research, and AI has begun to provide valuable support to researchers in this regard. Changes in the genetic composition of tumors translate into structural changes in cellular subsystems that need to be integrated into drug design to predict therapeutic response and concurrently learn about the mechanism underlying a particular drug response. A proper understanding of the mechanism of drug action can lead researchers to understand the importance of different signaling pathways, including some new and uncommon pathways associated with tumors, to help develop novel drugs for the therapeutic targeting of diverse forms of cancer. Drug combinations targeting multiple pathways are thought to address the incidence of drug resistance in cancer therapy, and computational models could be used to find solutions [112,113].

## 8. Advences in precision oncology and targeted drug therapy of cancer

Targeted cancer therapy is a form of cancer therapeutic that targets specific genes and proteins involved in cancer cell reprogramming, signaling molecules, and other molecules in the tumor microenvironment that contribute to cancer development. This contrasts with the single-target approach employed in chemotherapy to primarily target and kill actively dividing cancer cells with serious side effects; thus, the emergence of targeted drug therapy can be seen as a natural outcome of decades of studies on the molecular reprogramming of affected cells in different cancers [114]. Some notable breakthroughs have been made in certain cancers, as a renewed understanding of the signaling pathways underlying cancer development has led to the development of specific molecular targeted drugs in past decades. For example, tamoxifen is a wonder drug in medical oncology approved by the FDA in 1977 for the management of estrogen receptor-positive (ER-positive or ER+) breast cancer and can be used to treat all stages of breast cancer and as adjuvant treatment to alleviate the after-effects of surgery and radiotherapy. Studies have long supported the role of hormones, particularly estrogen, in the pathogenesis of breast cancer. Tamoxifen is essentially a hormone therapy drug that acts as an estrogen receptor antagonist to minimize the growth of breast cancer cells. It is among the first discovered to selectively target cancer cells with far fewer side effects and has successfully saved lives for decades and revolutionized the field of targeted cancer therapies [115]. This form of cancer therapy can be thoroughly optimized by means of precision oncology, which enables the use of genomic profiling of patient samples for insights into the mutational changes underlying pathway alterations responsible for cancer initiation and progression. Thus, precision oncology-based treatment strategies pledge the diagnosis and prognosis of cancer via the use of specific molecular-level information about a patient's tumor to treat the illness with selective targeting of affected cells with the desired results. In this way, this method can also be considered as a perfect theranostic approach for cancer treatment. The term ‘theranostic’ literally means a combination of diagnosis and therapeutics and refers to the pairing of diagnostic methods such as the proteogenomic approach to biomarker discovery, with appropriate therapeutic interventions for effective management of the disease. Theranostics focuses on patient-centered care and thus provides a transition from conventional to personalized medicine for targeted, efficient and safe pharmacotherapy relevantly applicable in precision oncology [116].

Cancer drugs employed in targeted therapies are primarily designed to target proteins directly involved in controlling the growth, division and spread of cancer cells. These are the intermediates of the signaling cascade of cancer cells or molecules of the tumor microenvironment essential for tumor growth and cancer manifestation. [117]. They are broadly classified as monoclonal antibodies (mAbs) or small-molecule drugs. Small- molecule drugs are designed to directly approach the cell membrane and interact with targets inside the cell and usually inhibit the enzymatic activity of target proteins such as the proteasome complex, cyclin-dependent kinases and a variety of signaling proteins. Kinase family proteins, such as tyrosine kinases, Rho kinases, Bruton tyrosine kinases, ABL kinases, and NAK kinases, play essential roles in modulating signaling pathways associated with cancer progression and therefore constitute valuable sources of anticancer agents against actionable targets in cancer therapeutics (Table 1).

Therapeutic targeting of DNA damage response (DDR) signaling is an emerging field of targeted cancer therapy. DDR signaling is essential for genome stability, and alteration in this signaling pathway is implicated in cancer progression. It also allows therapeutic options as cells with excessively defective DDR signaling are directed toward an alternative pathway that includes the induction of immunoregulatory signaling, apoptosis or senescence. These vulnerabilities have been exploited for anticancer treatments.

**Table 1.** List of Protein Kinase Inhibitors approved by FDA.

(NRY, nonreceptor protein-tyrosine kinase; RTK, receptor protein-tyrosine kinase; S/T protein-serine/threonine kinase; T/Y, dual-specificity protein kinase)

| Protein kinase inhibitor | Approval year | Primary targets | Target kinase family | Indications |
|---|---|---|---|---|
| Abemaciclib | 2017 | CDK4/6 | S/T | Breast cancer |
| Acalabrutinib | 2017 | BTK | NRY | Lymphoma |
| Afatinib | 2013 | ErbB1/2/4 | RTK | Lungcancer |
| Alectinib | 2015 | ALK, RET | RTK | Lungcancer |
| Avapritinib | 2020 | PDGFR | RTK | Gastrointestinal cancer |
| Axitinib | 2012 | VEGFR1/2/3 | RTK | Kidney cancer |
| Binimetinib | 2018 | MEK1/2 | T/Y | Melanoma |
| Bosutinib | 2012 | BCR-Abl | NRY | Leukemia |
| Brigatinib | 2017 | ALK | RTK | Lungcancer |
| Cabozantinib | 2012 | RET, VEGFR2 | RTK | Thyroid, kidney, Hepatocellular cancer |
| Capmatinib hydrochloride | 2020 | c-MET | RTK | Lungcancer |
| Ceritinib | 2014 | ALK | RTK | Lungcancer |
| Cobimetinib | 2015 | MEK1/2 | T/Y | Melanoma |
| Crizotinib | 2011 | ALK, ROS1 | RTK | Lungcancer |
| Dabrafenib | 2013 | B-Raf | S/T | Melanoma, lung, thyroid Cancer |
| Dacomitinib | 2018 | EGFR | RTK | Lungcancer |
| Dasatinib | 2006 | BCR-Abl | NRY | Leukemia |
| Encorafenib | 2018 | B-Raf | S/T | Melanoma, colorectal cancer |
| Entrectinib | 2019 | TRKA/B/C, ROS1 | RTK | Lungcancer; solid tumors |
| Erdafitinib | 2019 | FGFR1/2/3/4 | RTK | Urothelial carcinoma |
| Erlotinib hydrochloride | 2004 | EGFR | RTK | Lung, Pancreatic cancer |
| Everolimus | 2009 | FKBP12/mTOR | S/T | Breast, kidney cancer, Neuroendocrine tumors |
| Fedratinib | 2019 | JAK2 | NRY | Myelofibrosis |
| Futibatinib | 2022 | FGFR2 | RTK | Cholangiocarcinomas |
| Gefitinib | 2003 | EGFR | RTK | Lungcancer |
| Gilteritinib | 2018 | Flt3 | RTK | Leukemia |
| Ibrutinib | 2013 | BTK | NRY | Lymphoma |
| Imatinib mesylate | 2001 | BCR-Abl | NRY | Leukemia; Gastrointestinal cancer |
| Infigratinib | 2021 | FGFRs | RTK | Cholangiocarcinoma |
| Lapatinib ditosylate | 2007 | ErbB1/2/HER2 | RTK | Breast cancer |
| Larotrectinib | 2018 | TRKA/B/C | RTK | Solid tumors |
| Lenvatinib | 2015 | VEGFR, RET | RTK | Hepatocellular, endometrial, Thyroid, Kidney cancer |
| Lorlatinib | 2018 | ALK | RTK | Lungcancer |
| Midostaurin | 2017 | Flt3 | RTK | Leukemia |
| Mobocertinib | 2021 | EGFRwith exon 20insertions | RTK | Lungcancer |
| Neratinib | 2017 | ErbB2/HER2 | RTK | Breast cancer |
| Nilotinib | 2007 | BCR-Abl | NRY | Leukemia |
| Osimertinib | 2015 | EGFRT790M | RTK | Lungcancer |
| Pacritinib | 2022 | JAK2 | RTK | Myelofibrosis |
| Palbociclib | 2015 | CDK4/6 | S/T | Breast cancer |
| Pazopanib hydrochloride | 2009 | VEGFR1/2/3 | RTK | Kidney cancer, soft Tissue sarcoma |
| Pemigatinib | 2020 | FGFR2 | RTK | Cholangiocarcinoma |
| Pexidartinib | 2019 | CSF1R | RTK | Tenosynovial giant cell tumor |
| Pirtobrutinib | 2023 | BTK | NRY | Lymphoma |
| Ponatinib hydrochloride | 2012 | BCR-Abl | NRY | Leukemia |
| Pralsetinib | 2020 | RET | RTK | Lungcancer |
| Quizartinib | 2023 | FLT3/STK1 | RTK | Leukemia |
| Regorafenib | 2012 | VEGFR1/2/3 | RTK | Gastrointestinal, Colorectal, Hepatocellular cancer |
| Ribociclib | 2017 | CDK4/6 | S/T | Breast cancer |
| Ripretinib | 2020 | KIT/PDGFR | RTK | Gastrointestinal cancer |
| Ruxolitinib phosphate | 2011 | JAK1/2/3, Tyk | NRY | Myelofibrosis |
| Selpercatinib | 2020 | RET | RTK | Lung, thyroid cancer |
| Selumetinib | 2020 | MEK1/2 | T/Y | Neurofibroma |
| Sorafenib tosylate | 2005 | VEGFR1/2/3 | RTK | Thyroid, Kidney, Hepatocellular cancer |
| Sunitinib malate | 2006 | VEGFR2 | RTK | Gastrointestinal, kidney, Pancreatic cancer |
| Temsirolimus | 2007 | FKBP12/mTOR | S/T | Kidney cancer |
| Tepotinib | 2021 | Met | RTK | Lungcancer |
| Tivozanib | 2021 | VEGFR2 | RTK | Kidney cancer |
| Trametinib | 2013 | MEK1/2 | T/Y | Melanoma |
| Trilaciclib | 2021 | CDK4/6 | S/T | Lungcancer |
| Tucatinib | 2020 | ErbB2/HER2 | RTK | Breast cancer |
| Vandetanib | 2011 | VEGFR2 | RTK | Thyroid cancer |
| Vemurafenib | 2011 | B-Raf | S/T | Melanoma, histiocytic sarcoma |
| Zanubrutinib | 2019 | BTK | NRY | Lymphoma |

Poly (ADP–ribose) polymerase (PARP) and inhibitors of poly (ADP–ribose) glycohydrolase (PARG) are the most important DNA repair enzymes that work synergistically in many different DDR pathways, including base excision repair, nonhomologous end joining, nucleotide excision repair, homologous recombination (HR), maintenance of replication fork stability and nucleosome remodeling. These enzymes are essentially involved in the process of single-strand break (SSB) repair, whose failure leads to the conversion of SSB into double-strand breaks (DSBs), which require repair by HR to prevent cell death. Such lethal genetic interactions, known as synthetic lethality, can be exploited to develop anticancer therapeutics, and the enzymes involved in DDR signaling fit the needs well [118]. Cancers with somatic and germline mutations in BRCA1/2 and other HR genes, such as ATM, ATR, etc., include pancreatic, prostate, breast, ovarian, and oral cancers, and inhibition of PARP activity may be an effective therapeutic strategy. PARP and PARG inhibitors have shown improved results in different forms of tumors and other potential targets are under investigation for safe use in combination therapy [119].

A type of targeted therapy, called tumor-agnostic therapy, uses anticancer agents to target cancer-specific alterations to treat the problem without requiring a focus on the cancer type or where the disease may have started in the body. Thus, the same drug can be used to treat different cancers if they share a particular genetic mutation or biomarker. This method allows for a more personalized treatment approach, as different individuals with the same genetic alteration can benefit from the same drug, regardless of their cancer type. Therapeutic mAbs are modified monoclonal antibodies that target antigens found on cancer cells or cytotoxic T lymphocytes in targeted cancer therapy. mAbs are important in cancer treatment because they may be exploited for potentiating the natural immune system by successfully utilizing changes in the immunogenicity of affected cells during oncogenesis. mAbs may be designed to coat cancer cells to be opsonized and destroyed by immune cells, block the activity of different cancer-specific antigens called neoantigens generated by cancer cells, or inhibit the activities of immune checkpoint proteins that promote immune evasion during cancer development [120,121]. Several immune checkpoint proteins are expressed by immune cells, such as T cells and cancer cells, which are capable of binding with other partner proteins to help cancer cells escape immune responses. Their activation limits vital immune cell activities such as T- cell infiltration and other effector cell functions, resulting in tumor formation. CTLA-4 is a checkpoint protein present on the T-cell surface that binds to another protein called B7, preventing T cells from killing other target cells, including cancer cells. Certain mAbs, also called anti-CTLA4 monoclonal antibodies, are used to block CTLA-4 and are widely used as immune checkpoint inhibitors in a variety of human cancers. Different forms of monoclonal antibody-based therapy have proven to be efficacious in cancer treatment and are becoming increasingly important tools in targeted cancer therapy [122,123]. Importantly, cancer cells express a number of protein antigens that can be recognized by cytotoxic T lymphocyte (CTL) T cells, thus providing a means for CTL-mediated cancer therapy. The targeting of transformed cells by CTLs may be crucial for the prevention of both hematological and solid tumors, and the roles of these cells are being explored in cancer immunotherapy (Fig. 2).

T-cell transfer therapy, also called adaptive immunotherapy or immune cell therapy, is a new form of cancer treatment designed to exploit the enhanced antitumour immune response of tumor antigen-specific CTLs found in tumors and has been used effectively against neoantigen-possessing cells in recent years. Two types of T-cell transfer therapy, tumor-infiltrating lymphocyte or TIL therapy and CAR-T-cell therapy have been used, and both involve harvesting autologous T cells infiltrated into the tumor, growing large number of these cells in vitro, and administering them to the patient for the desired results. CAR-T-cell therapy is similar to TIL therapy except that T cells are designed to express a type of protein known as CAR (CAR for chimeric antigen receptor) to target specific antigens expressed in cancer cells in the body. Although CAR-T cells have significantly improved the treatment landscape of hematological malignancies, they have shown limited results in solid tumors, as solid tumors present certain obvious barriers to adoptive T-cell transfer and localization, but a variety of approaches are being developed to overcome these barriers to increase their specificity, efficacy, and safety in the treatment of different malignancies. Importantly, CAR-T therapy for hematological tumors typically attacks a single tumor target. Solid tumors like glioblastoma typically have a heterogeneous population of tumor cells, suggesting that treatments require multiple targets to be successful. Researchers recently tested a dual CAR-T therapy to try to overcome the immune defenses of glioblastoma, where the treatment is injected directly into the cerebrospinal fluid, and the results are encouraging [124]. The development of CAR-T-cell therapy for solid tumors has been impaired also because most target antigens are similar to those of normal cells. Research is being directed to develop a toolbox of novel chimeric antigen receptors (CARs) that could be programmed to use logic to discriminate between normal and cancerous cells to prevent toxicity. This development could help overcome some of the barriers to the application of CAR-T cells against solid tumors [125].

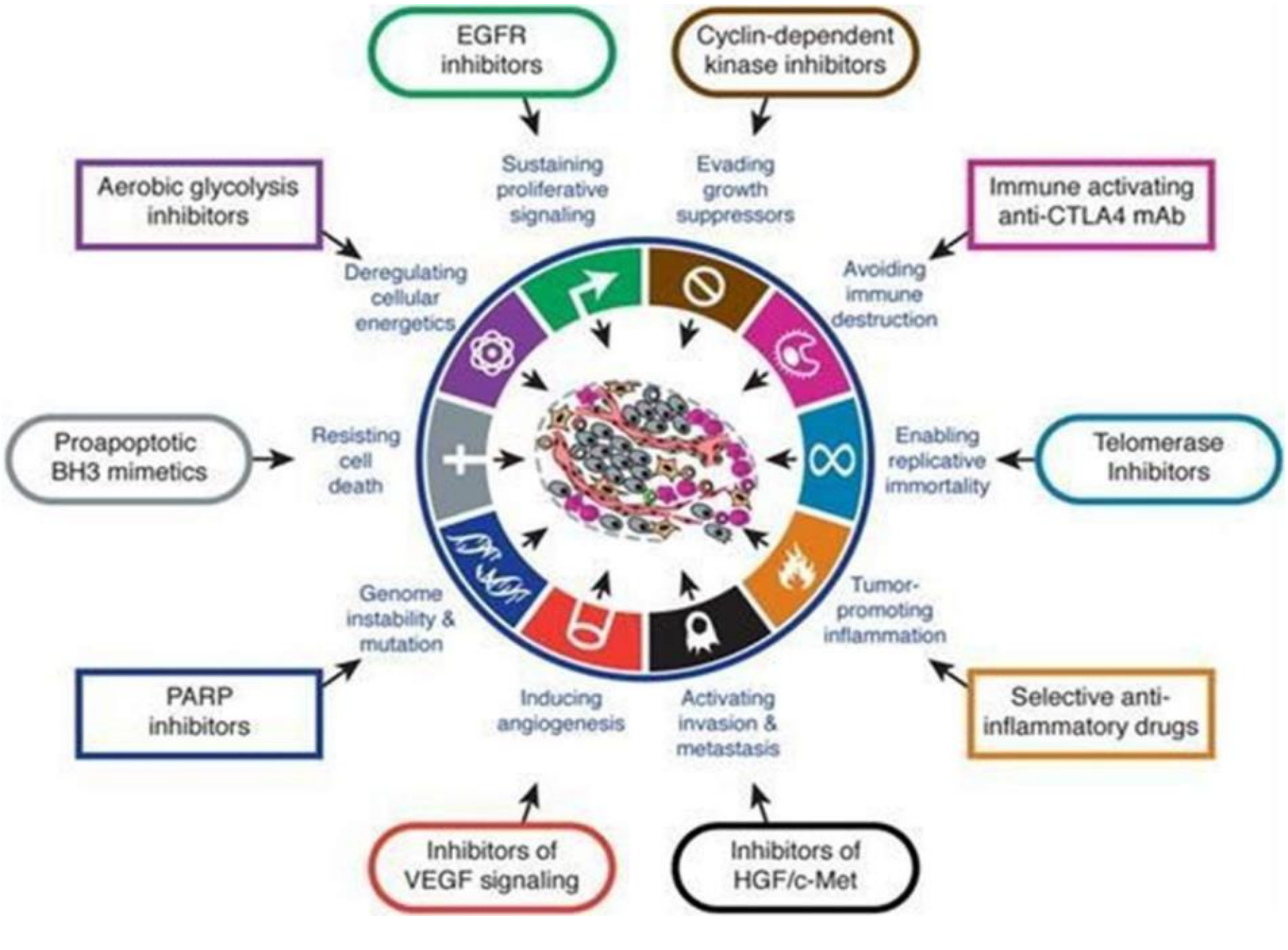

**Figure 2.** Therapeutic Targeting of the Hallmarks of Cancer
Therapeutic agents that can mitigate the acquired capabilities necessary for tumor growth and cancer progression are being developed for clinical use in treating different cancer types. These drugs are being developed in clinical trials to target each of the emerging neoplastic characteristics and the enabling hallmark capabilities for effective cancer therapy. The listed drugs are illustrative examples; there is a deep pipeline of investigational drugs in development to target different signaling molecules that lead to hallmark capabilities. (Hanahan and Wienberg [57]. With permission from Elsevier)

Furthermore, therapeutic cancer vaccines, such as dendritic cell (DC) vaccines,

peptide vaccines, and RNA-based neoantigen vaccines, have been developed to induce CTLs against antigens in cancer patients and have shown encouraging results. These vaccines can be designed to induce the production of biomolecules capable of targeting the shared antigens expressed by cancer cells through appropriate immune responses and are being investigated for their efficacy as neoantigen-targeted individualized cancer vaccines. Dendritic cells (DCs) are specialized antigen-presenting cells (APCs) known for their ability to present antigens to T cells, and this property of DCs has been exploited for their application in therapeutic cancer vaccines, which have been shown to induce protective antitumor activities. [126].

In addition, the transposable elements (TEs) usually present in the TME are potentially useful for creating a pancancer vaccine that can aid in the prevention of a range of cancers. An enumerable number of regions with TEs are involved in the expression of proteins in cancer cells. Many of these are shared across tumors of the same type and could provide means for destruction by the immune system. The ultimate goal of immunotherapy is to activate an individual's immune system against evolving tumors so that transformed cells can be successfully targeted with high selectivity, low toxicity, and appropriate outcomes.
Thus, cancer treatment will be driven by immunotherapy as a frontline area of cancer research, and precision oncology will rely on immunotherapy accordingly.

As discussed earlier, a major concern in cancer therapeutics is proper drug delivery to the affected cells and tissue for the desired outcomes. Conventional chemotherapeutics may have several serious side effects due to nonspecific targeting or inability to enter the core of the tumor, resulting in impaired treatment and a low survival rate. Researchers have been trying to address this issue with more specific methods of drug delivery, including the use of nanoparticles (NPs) in cancer therapeutics. NP-based systems can be programmed to recognize cancerous cells for selective and accurate drug delivery with increased drug localization, cellular uptake, and bioavailability, avoiding encounters with healthy cells. Newly developed quantum dots (QDs) are a class of heterogeneous fluorescent nanoparticles, the nanoscale materials with sizes ranging from 1 to 10 nm, unique optical properties and optimal surface chemical properties to link with targets such as antibodies, peptides, and other small-molecule drugs. As photoluminescent nanostructures possess fully quantized energy states with superior fluorescence characteristics, they are thought to be more specific and effective methods with wide applications in the diagnosis and molecular targeting of transformed cells. NP-based drug delivery systems, in general, display better pharmacokinetic and pharmacodynamic profiles, including efficient targeting of cancer cells and a reduction in side effects, and are sure to serve the needs of precision oncology-based therapy satisfactorily [127,128]. Furthermore, the use of antibody–drug conjugates (ADCs) is a fast-expanding therapeutic strategy designed to selectively deliver drugs to cancer cells. ADCs are monoclonal antibodies linked with small-molecule cytotoxic drugs through a chemical linker capable of approaching cancer cells and attaching to specific tumor antigens on the cell surface for direct drug delivery, sparing healthy cells in the surroundings. They are designed to exploit the features of antigen–antibody specificity for efficient drug delivery and are considered to be magic bullets in targeted cancer therapy [129, 130]. In this way, precision oncology seems to be the best fit for strategizing effective means of targeted drug therapy by exploiting the genomic peculiarities of individuals or a cohort of patients for effective personalized cancer treatment. Rigorous research on the genetic profile of cancer cells will continue to gain a thorough understanding of alterations in key signaling pathways and related molecular events during cancer progression, therapy resistance, and recurrence to help improve targeted cancer therapy.

Recent advances in cancer genomics and single-cell technologies have made targeted therapy the accepted form of cancer treatment; however, a large amount of investment will be needed for future research, drug discovery, and diagnostics to fully unlock its potential and for its application in the management of cancer. The socioeconomic burden of cancer remains high, as the treatment options for most common cancers have been limited thus far, which is an indication for a renewed approach to expedite drug development to bring effective anticancer agents from the bench to the bedside in a cost-effective manner. The lack of understanding of the genetic heterogeneity of individual cancers has traditionally limited the search for efficacious agents for cancer treatment, and a wide range of possibly suitable agents from other disease areas has been missed. The use of molecular characterization of different cancer types through cancer genomics can help resolve drug- related issues to a reasonable extent by repurposing certain existing drugs as anticancer agents for a wide range of applications, and it will remain at the forefront of precision oncology [131,132]. Furthermore, the move from tissue-based cancer-specific treatments to genome-based targeted treatments entails the reuse of anticancer drugs prescribed for one type of cancer to treat other cancer types as well. With the increasing understanding of cell signaling mechanisms and genetic alterations in carcinogenesis, considerable progress in cancer treatment may be realized in the near future. Continued research in this area holds the opportunity for improving cancer diagnosis, prevention, and treatment, thereby improving results.Considering that academia, industries, and civil society will work in tandem to cater to the contemporary needs of the system, it is hoped that a wide range of people with cancer will benefit from this new development in cancer research in the future to benefit the system as a whole [133,134].

## 9. Conclusion

The advancements in cancer genetics have profoundly influenced our understanding of cancer biology and the development of targeted therapies. Precision oncology-based cancer therapeutics proposes the development of treatments that target the specific molecular characteristics of an individual's tumor instead of targeting the common features of certain cancers for a cure. Considering that a thorough understanding of the genetic composition and heterogeneity of an individual's tumor is now becoming possible through single-cell technologies, it is poised to help individuals obtain the right treatment at the right time rather successfully without requiring more generalized treatment that would ultimately prove ineffective. Furthermore, cancer research has traditionally focused on common cancers for obvious reasons, leaving therapeutic options for less frequent tumor types largely limited, and such anomalies are likely to be successfully addressed with new developments. In addition, precision medicine approaches to treat inherited diseases have been used for directly targeting associated pathways and proteins, and such methods can also be employed in the treatment of inherited cancers. Importantly, drug resistance has traditionally been a serious problem in cancer treatment, but the emergence of targeted drug therapy based on precision oncology can greatly improve outcomes. The evolution of gene detection methods, liquid biopsy, and single-cell sequencing technology could facilitate deciphering the molecular mechanism of tumor drug resistance to help develop updated and effective anticancer agents in response to drug resistance. Thus, precision oncology, which relies on the genomic specificity of individuals for successful targeting of the most specific pathways involved in disease progression, is best suited to ensure precise treatment of the disease.

This is, in fact, a natural outcome of cancer genome research; the level of support from multiomics platforms is most encouraging, and it is poised to satisfactorily achieve the intended goal of cancer initiatives. The growing success of this form of treatment is sure to further strengthen our belief in the possibility of an effective treatment for cancer, and it must be made available to an increasing number of people with cancer to achieve the goals over time.

**Acknowledgements:** I acknowledge the continued support of the School of Bio Sciences & Technology, Vellore Institute of Technology, Vellore, Tamil Nadu, India, which has enabled this work to be completed over time.

**Authors' contributions:** Not applicable

**Competing interests:** Not applicable

**Informed consent statement:** Not applicable

**Funding:** Not applicable